\documentclass{IEEEtran}
\usepackage{cite}
\usepackage{amsmath,amssymb,amsfonts}
\usepackage{algorithmic}
\usepackage{graphicx}
\usepackage{tikz}
\usepackage{subcaption}
\usepackage{textcomp}
\usepackage{bm}
\usepackage{dutchcal} 
\usepackage[colorlinks=true,citecolor=blue,linkcolor=blue,urlcolor=blue]{hyperref}
\def\BibTeX{{\rm B\kern-.05em{\sc i\kern-.025em b}\kern-.08em
		T\kern-.1667em\lower.7ex\hbox{E}\kern-.125emX}}
\begin{document}
	\title{
		Angular-Channel Capacity and Nonlocal Wavefront Engineering for Phased-Array–Lens Systems
	}
	\author{Mohammad Soltani, \IEEEmembership{Graduate Student Member, IEEE}, and George V. Eleftheriades, \IEEEmembership{Fellow, IEEE}
			\thanks{This work was supported by the Natural Sciences and Engineering Research Council (NSERC) of Canada.}
			\thanks{The authors are with the Edward S. Rogers Sr. Department of Electrical and Computer Engineering, University of Toronto, Toronto, Ontario, M5S 3G4, Canada. (e-mail: m.soltani@mail.utoronto.ca, gelefth@waves.utoronto.ca)}
	}
	
	\maketitle
	
	\begin{abstract}
		Wide-angle beam steering with high directivity is fundamentally limited by diffraction, restricting the scan performance of phased-array systems. Here, we establish the bounds governing beam steering with high directivity in phased-array–lens systems and show that nonlocal dielectric metalenses can operate close to these limits. Unlike conventional metalenses that suppress mutual coupling, our approach exploits nonlocal electromagnetic interactions to laterally redistribute energy and expand the spatial width of phase coherence across the aperture. This enables efficient wavefront engineering for uniformly excited planar phased arrays using conventional linear progressive phasing. We further reveal the roles of diffraction and evanescent waves in determining the maximum coherent-aperture expansion and directivity enhancement. The resulting angular-channel bounds identify two distinct operating regimes for passive linear phased-array–lens systems: scan-resolution enhancement and directivity enhancement with preserved scan range. Guided by these limits, we derive geometry-independent design equations and develop an adjoint-based inverse-design framework for passive linear metalenses, enabling planar and cylindrical implementations that operate close to the fundamental limits. We experimentally validate a compact cylindrical metalens that preserves the $50^\circ$ scan range and scan resolution of a phased array with a $3.5\lambda_0$-wide aperture while providing a $2.5$--$3.5\,\mathrm{dB}$ increase in peak directivity. These results establish nonlocal dielectric metalenses as a practical platform for diffraction-limited wavefront engineering in phased arrays.
	\end{abstract}
	
	
	\section{Introduction} \label{sec:introduction}
	
	Phased arrays are indispensable in modern wireless communications and object detection and ranging systems \cite{Heterogeneously_Integrated,Mailloux,Rebeiz-1024}. Their performance is fundamentally tied to the size of the effective aperture, which largely determines the achievable beam quality and directivity. Increasing the aperture, however, typically requires a large number of radiating elements. As a result, increasing directivity can substantially increase system cost and complexity. This challenge becomes particularly acute at millimeter-wave and higher frequencies, where rapidly increasing free-space propagation loss makes these systems increasingly gain-limited rather than bandwidth-limited \cite{Quevedo-Teruel-5G}. Similarly, wide-angle optical phased arrays are emerging as key technologies for applications including light detection and ranging, free-space optical communications, and quantum photonics \cite{Hajimiri, Notaros}. In these systems, aperture size likewise governs beam quality, directivity, and free-space coupling efficiency, while scaling the aperture is constrained by photolithography reticle limits and packaging considerations. Consequently, improving directivity is often prohibitively expensive: a 3-dB increase in directivity requires doubling the aperture size and, therefore, approximately doubling the total number of radiating elements.
	
	Recently, nonlocal metasurfaces have emerged as a means of extending wave transformations beyond the local, position-dependent response of conventional metasurfaces, enabling angle-dependent and spatially distributed electromagnetic transformations \cite{Shastri-NonlocalFlatOptics}. More broadly, multifunctional metasurfaces can independently transform incident beams distinguished by their spatial, spectral, or polarization characteristics into corresponding output beams with tailored spatial or spectral responses. Such metasurfaces provide a large number of design degrees of freedom, enabling sophisticated field transformations for applications including antenna beamforming \cite{Vasilis-bandwidth, Szymanski-inverse,Vasilis-Modulated} and analog signal processing \cite{Estakhri, Momeni-beamforming}. To realize these functionalities, gradient-based inverse-design methods with analytically computed sensitivities have emerged as a powerful framework for solving large-scale electromagnetic optimization problems \cite{Hammond-inverse, ZinLin-inverse, Ji-Adjoint, Zhaoyi-Empowering}. Notably, inverse design has enabled metalenses to compensate for chromatic \cite{Zin-Lin-multiwavelength, Zhaoyi-Achromat} and Seidel \cite{Shiyu-metalens} aberrations beyond what is achievable using conventional design approaches. Despite these advances, the underlying physical limits governing multifunctional wave transformations remain poorly understood. In particular, there is currently no rigorous physical framework explaining the reduction in scan range observed when a conventional hyperbolic phase-delay lens is combined with a phased array \cite{Tamijani}, nor are the corresponding fundamental performance limits known.
	
	This work establishes the physical principles governing passive, linear metalenses for phased-array antennas. Specifically, we show how diffraction fundamentally limits the achievable coherent-aperture expansion and, consequently, the directivity enhancement attainable with passive lenses. We further identify the critical role of evanescent-wave-assisted energy transport in approaching these limits. To establish this framework, we consider metalenses that transform an incident field with a planar wavefront and coherent-aperture width $W_{\mathrm{in}}$ into a transmitted field that retains the planar wavefront while exhibiting a larger coherent-aperture width, $W_{\mathrm{out}}>W_{\mathrm{in}}$. In other words, the coherent output aperture is the portion of the transmitted field that behaves as a single enlarged antenna aperture. Increasing this width allows fields over a larger aperture to interfere constructively, producing a narrower radiation beam and higher directivity. Building on this physical framework, we derive design equations for two classes of passive, linear metalenses: scan-resolution-enhancing metalenses, which improve scan resolution through angular-channel compression, and scan-range-preserving metalenses, which enlarge the coherent aperture while maintaining the available field of view.
	
	To realize these concepts, we develop an adjoint-based inverse-design framework formulated with integral equations, enabling efficient optimization while rigorously accounting for full-wave electromagnetic interactions, including multiple scattering and mutual coupling. Unlike conventional optical metalenses, which rely on local phase control for beam focusing, the proposed metalenses are designed to laterally redistribute electromagnetic energy through engineered nonlocal interactions, allowing evanescent-wave and near-field coupling to expand the coherent-aperture width of the transmitted field. Rather than suppressing mutual coupling, our approach exploits it as the underlying mechanism for wavefront synthesis, enabling operation with conventional phased arrays employing standard element spacing, uniform excitation amplitudes, and progressive phase shifts. The methodology is demonstrated for both planar and cylindrical metalenses using parametrized dielectric meta-atoms that balance computational efficiency, fabrication simplicity, and design flexibility. The resulting multilayer dielectric structures can be fabricated using low-cost additive manufacturing. Through representative wide-angle designs, we demonstrate that cylindrical metalenses provide a compact alternative to planar implementations while preserving scan performance. Together, these results establish practical design principles for exploiting, rather than suppressing, mutual electromagnetic coupling, enabling compact, manufacturable, and high-efficiency dielectric metalenses that operate close to the fundamental diffraction limits while substantially improving phased-array performance over broad bandwidths and wide scan angles.
	
	The paper is organized as follows. Section~\ref{Sec:Fundamental_Limits} establishes the physical framework for phased-array--lens systems and identifies two distinct operating regimes. Section~\ref{Sec:Analysis_and_Synthesis} then presents the method-of-moments (MoM) analysis framework and the metalens synthesis procedure based on gradient-descent optimization using the adjoint method. Section~\ref{Sec:Design_Platform} describes the constitutive unit cell and the nonlocal coupling mechanism used to realize the proposed wavefront transformations. Building on these foundations, Sec.~\ref{Sec:Phased_Array_Performance_Enhancement} demonstrates the enhancement of phased-array performance through two physically realizable metalens designs: a scan-resolution-enhancing metalens in Sec.~\ref{SubSec:Scan_Resolution} and a scan-range-preserving metalens in Sec.~\ref{SubSec:Scan_Range}. The latter is further validated experimentally through measurements of a coherent-aperture-expanding cylindrical metalens prototype. Finally, the conclusions are presented in Sec.~\ref{Sec:Conclusion}.	
	
	\begin{figure}[!t]
		\centerline{\includegraphics[width=\columnwidth]{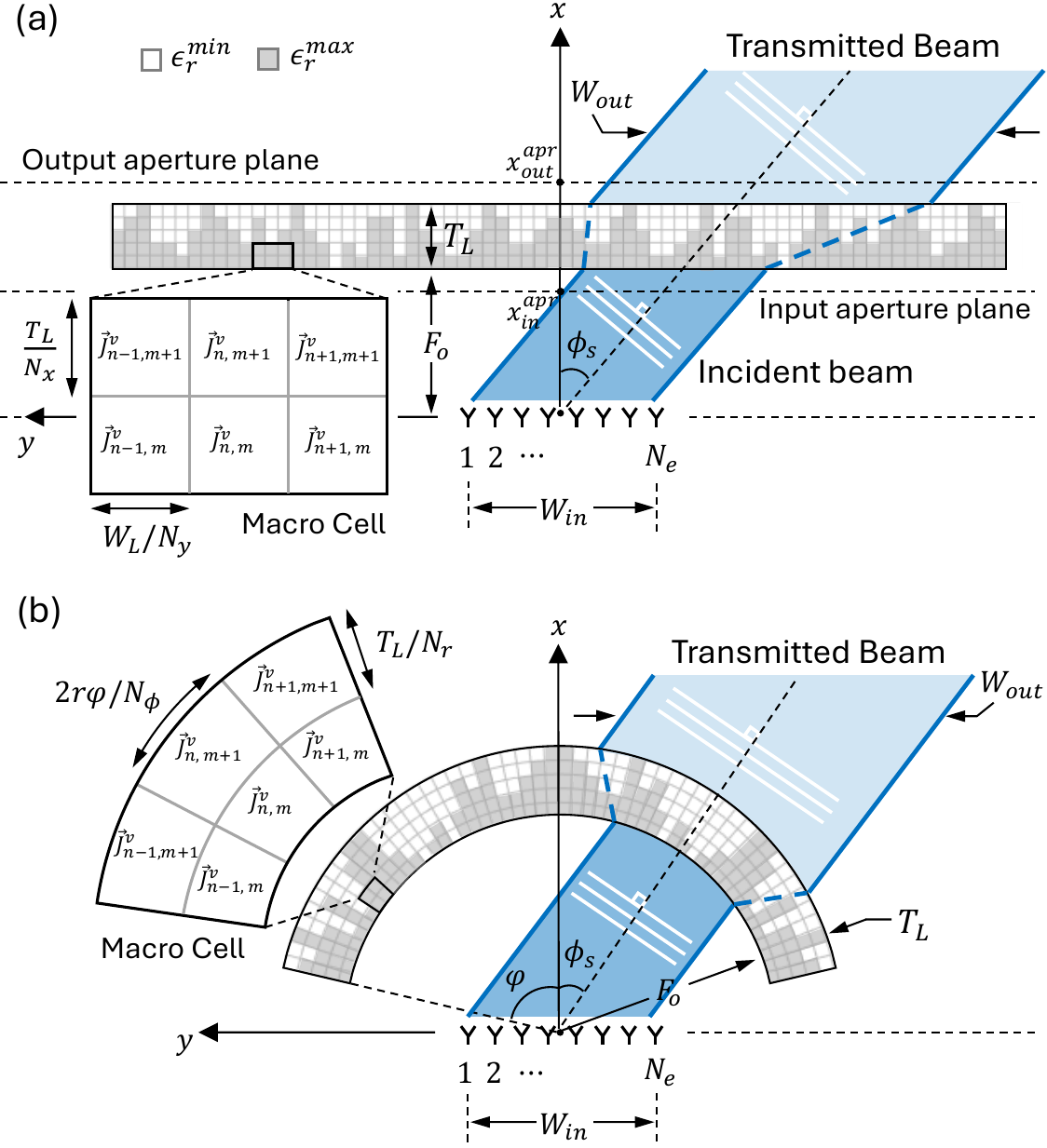}}
		\caption{Geometry of the (a) flat, (b) cylindrical shell metalens and the phased array.}
		\label{fig:Geometry}
	\end{figure}
	
	\section{Angular-Channel Capacity in Phased-Array–Lens Systems} \label{Sec:Fundamental_Limits}
	Consider a linear, time-invariant, and passive lens placed in front of a phased array (PA), as shown in Fig.~\ref{fig:Geometry}. The array comprises $N_e$ elements with uniform spacing $d_e$, giving an input aperture of width $W_{\mathrm{in}}=(N_e-1)d_e$. We restrict the analysis to the propagating spectrum of the array field. Accordingly, the array is excited such that its radiated field is dominated by plane-wave components satisfying $|k_y|\leq k_0=2\pi/\lambda_0$, where $k_0$ and $\lambda_0$ denote the free-space wavenumber and wavelength, respectively.
	
	Because the radiated field is confined both spatially to an aperture of width $W_{\mathrm{in}}$ and in spatial frequency to a bandwidth $2k_0$, the propagating-mode subspace $\mathcal K_{\mathrm{in}}$ has an approximate dimension set by the space–bandwidth product \cite{Landau-Pollak-III}:
	\begin{equation} \label{SBP_Dimension}
		\dim(\mathcal{K}_{\mathrm{in}}) = N_{\mathrm{in}} \approx \frac{2W_{\mathrm{in}}}{\lambda_0}.
	\end{equation}
	Physically, this means that although the field is a continuous function, only approximately $N_{\mathrm{in}}$ independent spatial modes can be supported within the specified aperture and propagating spectrum. The modes are not separated by a perfectly sharp cutoff; rather, the dominant modes contain most of the supported field content, while the modal strength decreases rapidly beyond the space–bandwidth limit. Thus, $N_{\mathrm{in}}$ provides an estimate of the maximum number of independent angular channels supported by the phased-array aperture. The corresponding minimum spatial-frequency resolution is
	\begin{equation}
		\Delta k_y^{\mathrm{in}} = \frac{2\pi}{W_{\mathrm{in}}}.
	\end{equation}
	This resolution is attained for uniform aperture excitation, whereas a smooth amplitude taper broadens the beamwidth \cite{Hansen}, thereby increasing $\Delta k_y^{\mathrm{in}}$ and degrading the achievable scan resolution.
	
	Ideally, a lens used to enhance the performance of a phased array transforms the propagating modes supported by the input aperture into the corresponding modes of a larger output aperture. A lens made of a linear material therefore defines a physical linear transformation between the input and output field spaces. Restricting this transformation to the propagating modal subspaces of $\mathcal K_{\mathrm{in}}$ and $\mathcal K_{\mathrm{out}}$ and representing it in the optimal bases yields a linear angular-channel transmission matrix $\mathbf T \in \mathbb C^{N_{\mathrm{out}}\times N_{\mathrm{in}}}$. Here, $N_{\mathrm{in}}$ and $N_{\mathrm{out}}$ are the dimensions of the input and output modal subspaces, respectively, with $N_{\mathrm{out}}=\dim(\mathcal K_{\mathrm{out}})\approx2W_{\mathrm{out}}/\lambda_0$.
	
	For this linear transformation in passive media, the corresponding operator satisfies
	\begin{equation}
		\operatorname{rank}(\mathbf T)
		\le
		\min
		\left(
		N_{\mathrm{in}},\,
		N_{\mathrm{out}}
		\right),
		\label{Lens_Rank}
	\end{equation}
	which establishes an upper bound on the number of independent propagating channels that can be transmitted by a passive linear lens. Within the truncated modal representation, this rank bound is exact. For the underlying continuous fields, this finite-dimensional bound represents the transition between dominant and weakly concentrated modes rather than an idealized hard cutoff \cite{Landau-Pollak-II}. Accordingly, the number of effectively independent output channels remains bounded by $\min(N_{\mathrm{in}},\, N_{\mathrm{out}})$. Increasing the width of the coherent aperture using a passive, linear lens does not create additional degrees of freedom; it redistributes the existing channels over a larger spatial region.
	
\begin{figure}[!t]
		\centerline{\includegraphics[width=\columnwidth]{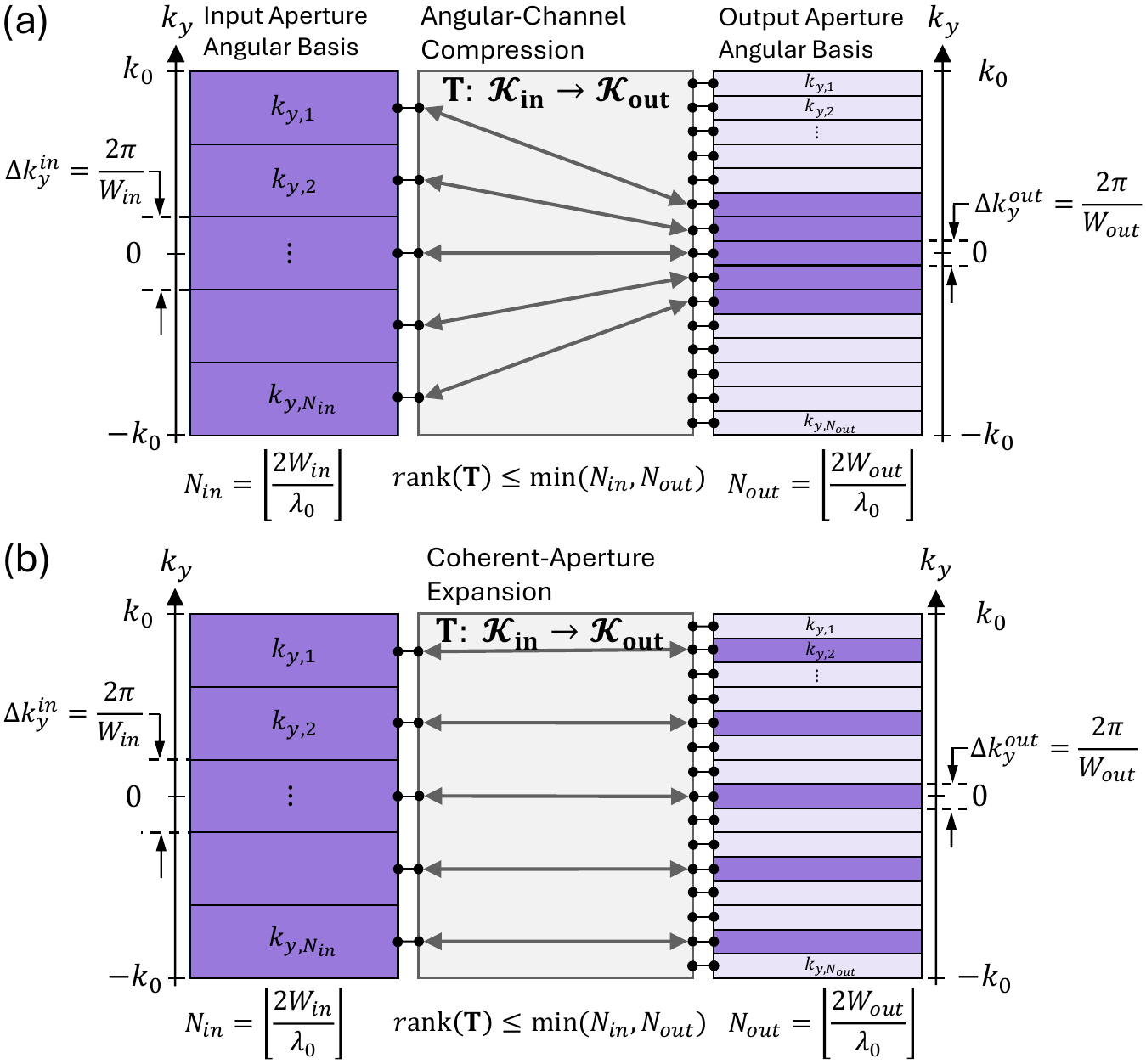}}
		\caption{Illustration of the fundamental bounds and the two operating regimes of the array–lens system, i.e., (a) angular-channel compression, and (b) coherent-aperture expansion.}
		\label{fig:Regimes}
	\end{figure}

	This channel-capacity constraint gives rise to two distinct operating regimes. In both regimes, the lens increases the peak directivity and narrows the transmitted beam. The regimes differ in how the available angular channels are distributed. In the first regime, which we term \emph{angular-channel compression}, the lens concentrates the available channels into a narrower angular range, as illustrated in Fig.~\ref{fig:Regimes}(a). Because their total number is fixed by Eq.~\eqref{Lens_Rank}, concentrating the channels into a smaller angular sector necessarily reduces the total angular span they occupy. The spacing between adjacent channels therefore decreases, providing finer angular sampling but over a reduced field of view. We refer to this channel spacing as the \emph{scan resolution}. Thus, the improvement in scan resolution is accompanied by a corresponding reduction in scan range.
	
	In the second regime, which we term \emph{coherent-aperture expansion}, the lens preserves both the field of view and the angular-channel spacing of the feeding array, as illustrated in Fig.~\ref{fig:Regimes}(b). The array therefore samples the same angular range with channels of higher directivity. Importantly, the increase in directivity is achieved without reducing the angular spacing between adjacent channels, allowing both the scan resolution and scan range to be preserved.
	
	\section{Analysis and Synthesis Framework} \label{Sec:Analysis_and_Synthesis}
	
	The phased-array–metalens configuration is shown in Fig. \ref{fig:Geometry}. The metalens is modeled as an inhomogeneous dielectric slab positioned in front of a phased array . The structure is assumed invariant along the $z$-axis, and the array currents are restricted to the $z$-direction. The radiated fields are computed using a finite set of line currents and the Green’s function of a grounded dielectric substrate, as described in \cite{soltani-anisotropic}.
	
	For the planar metalens configuration in Fig. \ref{fig:Geometry}(a), the lens is located at a distance $F_o$ from the array, with thickness $T_L$ and width $W_L$, discretized into $N_x \times N_y$ cells. For the cylindrical configuration in Fig. \ref{fig:Geometry}(b), the lens is positioned at a radial distance $F_o$ from the array center, spans an angular extent of $2\varphi$, and has thickness $T_L$; it is discretized into $N_r \times N_\phi$ cells in the $r$- and $\phi$-directions, respectively. 
	
	\subsection{Integral-Equation Formulation}
		We assume that the incident electromagnetic wave is transverse-electric (TE) polarized ($\bm{E}_{i} = E_{i} \, \hat{z}$), which is the dominant field component generated by the phased array when scanning in the H-plane. Together with the uniformity of the structure along the $z$-axis, this implies that the scattered field is also TE polarized, allowing the problem to be reduced to a two-dimensional cross-section. Finally, a time-harmonic dependence $e^{j \omega t}$ is assumed for all electromagnetic quantities throughout the manuscript.
	
		In this full-wave integral-equation formulation, the incident field induces a polarization current density $\bm{J} = J_z \hat{z}$ in the dielectric region $S_v$. This induced current produces a secondary scattered field $\bm{E}_s = E_s \hat{z}$. Since the dielectric is modeled with a polarization current, the corresponding scattered fields are calculated based on a two-dimensional free-space Green's function as
		\begin{equation} \label{Scattered_field}
			\bm{E}_{s}(\bm{r}) = - \frac{k_0 \eta_0}{4} \iint\limits_{Sv} H_0^{(2)} \left(k_0 |\bm{r} - \bm{r^\prime}|\right) \bm{J}(\bm{r^\prime}) ds^\prime,
		\end{equation}
		where $H_0^{(2)}(\cdot)$ is the zeroth-order Hankel function of the second kind, $\bm{r}$ and $\bm{r^\prime}$ are the position vectors of the observation and source points, respectively, and $\eta_0 = 120 \pi$ is the free-space impedance.
		
		The total electric field $\bm{E}$, given by the superposition of the incident field $\bm{E}_{i}$ and the scattered field $\bm{E}_{s}$, satisfies the constitutive relation in the dielectric region \cite{Harrington}:
		\begin{equation}\label{total_field}
			\begin{split} 
				\bm{E}(\bm{r}) &= \bm{E}_{i}(\bm{r}) + \bm{E}_{s}(\bm{r}) \\
				&= \frac{1}{j \omega (\epsilon_r (\bm{r}) - 1) \epsilon_0} \bm{J}(\bm{r}), \quad \text{in $S_v$,}
			\end{split}
		\end{equation}
		where $\omega$ is the angular frequency of the wave, $\epsilon_0 = 8.85 \times 10^{-12} $ F/m is the free-space permittivity, and $\epsilon_r (\bm{r})$ is the position-dependent dielectric constant. Equations \eqref{Scattered_field} and \eqref{total_field} form an integral equation for the unknown current density $\bm{J}(\bm{r})$, which is solved using the method of moments. Specifically, the volumetric current density is expanded in terms of non-overlapping two-dimensional pulse basis functions. The discretization grid is illustrated in Fig.~\ref{fig:Geometry}(a,b) for the flat and cylindrical structures, respectively.
		
		After expanding the induced current density using $N_v$ pulse basis functions and applying point matching at the center of each cell, the integral equation is reduced to the following linear system:
		\begin{equation} \label{Linear_System_of_Equation}
			\bm{E}_{i,v} = \left( \bm{X} - \bm{G}_{vv} \right)\bm{J}_v \triangleq \bm{L}\bm{J}_v,
		\end{equation}
		where $\bm{E}_{i,v}$ is the column vector of incident electric-field samples evaluated at the center of each discretization cell, $\bm{J}_v$ contains the unknown amplitudes of the pulse basis functions representing the induced current density, and $\bm{L}$ is the system matrix of the discretized integral equation. The diagonal matrix $\bm{X}$ describes the local material response and has dimensions $N_v\times N_v$, with diagonal entries
		\begin{equation}
			X_{nn} = \frac{1}{j\omega(\epsilon_r^n-1)\epsilon_0}.
		\end{equation}
		where $\epsilon_r^n$ denotes the relative permittivity of the $n$-th cell.
		
		Solving \eqref{Linear_System_of_Equation} determines the amplitudes of the induced current density $\bm{J}_v$ within $S_v$ for a specified set of cell dielectric constants $\epsilon_r^n$. The far-field $\bm{E}^{ff}(\phi)$ is obtained by superposing the contributions from all induced currents and the incident source, from which the radiation intensity,
		\begin{equation} \label{Radiation_Intensity}
			U(\phi) = \frac{1}{2 \eta_0} \left|\sqrt{r} \bm{E}^{ff}(\phi) \right|^{2},
		\end{equation}
		is obtained.
	
	\subsection{Gradient Descent Optimization Using the Adjoint Method}
	
		We define the cost function as the Euclidean norm of the difference between the desired and transmitted radiation intensity distributions over a prescribed angular interval centered at the scan angle $\phi_s$:
		\begin{equation}\label{Cost_Funtion}
			F = \left\lVert \Delta \bm{U} \right\rVert_{2},
		\end{equation}
		where
		\begin{equation}
			\Delta \bm{U} = \bm{U}_{d} - \bm{U}_{t},
		\end{equation}
		with $\bm{U}_{d}$ denoting a column vector containing samples of the radiation intensity of a uniformly excited aperture of width $W_{\mathrm{out}}$, with a linear excitation phase profile $k_0y\sin\phi_{t}$. The radiation intensity is sampled at $N_d$ uniformly spaced angular points over an angular interval centered at the desired beam direction $\phi_{t}$, with a total span of approximately five times the 3-dB beamwidth of the desired beam. This angular interval is selected because it captures the main lobe and the first two sidelobes while keeping the computational cost and matrix dimensions manageable. The vector $\bm{U}_{t}$ contains samples of the radiation intensity of the array–lens system at the same angular points.
		
		The far-field electric field can be obtained from the induced currents as
		\begin{equation} \label{Near-field_to_Far-field}
			\bm{E}^{ff} = \bm{E}^{ff}_i + \bm{R} \; \bm{J}_v,
		\end{equation}
		where $\bm{E}^{ff}_i$ is a column vector containing the far-field electric field of the source, and $\bm{R}$ is the radiation matrix, which contains the near-field-to-far-field current-to-electric-field transformation coefficients \cite{Harrington}, given by
		\begin{equation} \label{Farfield_Radiation_Equation}
			R_{mn} = -j \omega \mu_0 \frac{e ^{-j k_0 r}}{\sqrt{8j \pi k_0 r}} e^{j k_0 \hat{\bm{r}}_{m} \cdot \bm{r}^{\prime}_{n}} \Delta s_{n} ,
		\end{equation}
		where $\mu_0 = 4 \pi \times 10^{-7}$ H/m is the free-space permeability, $\hat{\bm{r}}_m$ is the unit vector in the direction of $\phi_m$, $\bm{r}_{n}^{\prime}$ and $\Delta s_n$ are the position vector and area of cell $n$, respectively. This operation is analogous to a Fourier transform of the near field, illustrating how spatial variations in the near field translate into angular variations in the far field.
		
		The gradient of the scalar cost function $F$ with respect to the design variable vector $\bm{p} = [p_1,\dots,p_N]^T$ is defined as
		\begin{equation}
			\nabla_{p} F =
			\left[
			\frac{\partial F}{\partial p_1},
			\cdots,
			\frac{\partial F}{\partial p_N}
			\right]^T.
		\end{equation}
		Using \eqref{Cost_Funtion}, the partial derivative of $F$ with respect to an individual parameter $p_n$ can be written as
		\begin{equation} \label{Variation_of_Cost_Function}
			\frac{\partial F}{\partial p_n} = \frac{-1}{\eta_0} \hat{\Delta\bm{U}}^{T} \; \Re \left\{ {\bm{E}^{ff}}^{*} \odot \frac{ \partial \bm{E}^{ff}}{\partial p_n} \right\},
		\end{equation}
		where we have used \eqref{Radiation_Intensity} to evaluate the derivative of $\bm{U}_M$ with respect to $p_n$. Here, $\hat{\Delta\bm{U}}$ is a unit vector in the direction of $\Delta\bm{U}$, and the operators $(\cdot)^T$, $\Re\{\cdot\}$, $\odot$, and $(\cdot)^*$ denote the transpose, real part, Hadamard (element-wise) product, and complex conjugation, respectively.
		
		From \eqref{Farfield_Radiation_Equation}, the derivative of the far-field vector with respect to $p_n$ is given by
		\begin{equation} \label{Variation_of_Eff_with_pk}
			\frac{\partial \bm{E}^{ff}}{\partial p_n} = \bm{R} \frac{\partial \bm{J}_{v}}{\partial p_n}.
		\end{equation}
		This expression highlights that the objective gradient is governed by the parametric sensitivity of the induced currents, propagated to the observation domain through the radiation operator.
		
		To compute $\partial \bm{J}_v / \partial p_n$, we differentiate the governing linear system \eqref{Linear_System_of_Equation} with respect to $p_n$, noting that the excitation $\bm{E}_{i,v}$ is independent of the design variables. Solving for the current sensitivity and substituting the result into \eqref{Variation_of_Eff_with_pk} gives
		
		\begin{equation} \label{Eff_variation_with_pk}
			\frac{ \partial \bm{E}^{ff}}{\partial p_n} = - \bm{R} \bm{L}^{-1} \frac{\partial \bm{L}}{\partial p_n} \bm{J}_{v}.
		\end{equation}
		Substituting in \eqref{Variation_of_Cost_Function}, we obtain
		\begin{equation}
			\frac{\partial F}{\partial p_n} = \frac{-1}{\eta_0} \hat{\Delta\bm{U}}^{T} \Re \left\{ {\bm{E}^{ff}}^{*} \odot {\bm{J}_{a}}^{T} \frac{\partial \bm{L}}{\partial p_n} \bm{J}_{v} \right\},
		\end{equation}
		where $\bm{J}_{a}$ is a matrix containing the amplitudes of the basis functions used to represent the polarization current density in the adjoint problem, and it satisfies the following linear system:
		\begin{equation}
			\bm{L}\, \bm{J}_{a} = \bm{E}_{i}^{a},
		\end{equation}
		where $\bm{E}_{i}^{a}$ is the adjoint excitation, given by
		\begin{equation} \label{Eia_expression}
			\bm{E}_{i}^{a} = - \bm{R}^{T}.
		\end{equation}
		This adjoint formulation expresses the objective gradient as a bilinear form involving the forward current $\bm{J}_{v}$, the adjoint current $\bm{J}_{a}$, and the parametric derivative of the system matrix. Importantly, the gradient can be computed with only two linear solves—one forward and one adjoint—irrespective of the number of design variables, thereby enabling efficient large-scale optimization. In this derivation, $\bm{L}^{-1}$ is symmetric. This follows from the reciprocity of the background medium, which gives $\bm{G}_{vv}^{T}=\bm{G}_{vv}$. Since $\bm{X}$ is diagonal and therefore satisfies $\bm{X}^{T}=\bm{X}$, the system matrix $\bm{L}=\bm{X}-\bm{G}_{vv}$ is symmetric, and its inverse inherits the same symmetry, i.e., $(\bm{L}^{-1})^{T}=\bm{L}^{-1}$.
		
		The expression in \eqref{Eia_expression} may be interpreted as the inverse Fourier transform of the far-field distribution associated with a plane wave incident at angle $\phi_m$. When this field is combined with contributions from plane waves spanning the angular interval centered at $\phi_{t}$, with each contribution weighted by the complex-conjugated far-field coefficient ${\bm{E}^{ff}}^{*}$, the resulting field distribution reconstructs the field of a rectangular aperture of width $W_{\mathrm{out}}$. In this framework, the term $\hat{\Delta \bm{U}}$ represents the normalized residual direction in the objective-function space, which determines the weighting of the adjoint sensitivity and, consequently, the descent direction in the design-parameter space.
	
	\section{Nonlocal Metalens Design Platform} \label{Sec:Design_Platform}
	\subsection{Constitutive Cell and Material Characterization} \label{Subsec:Cell_and_Material}
\begin{figure}[!t]
		\centerline{\includegraphics[width=\columnwidth]{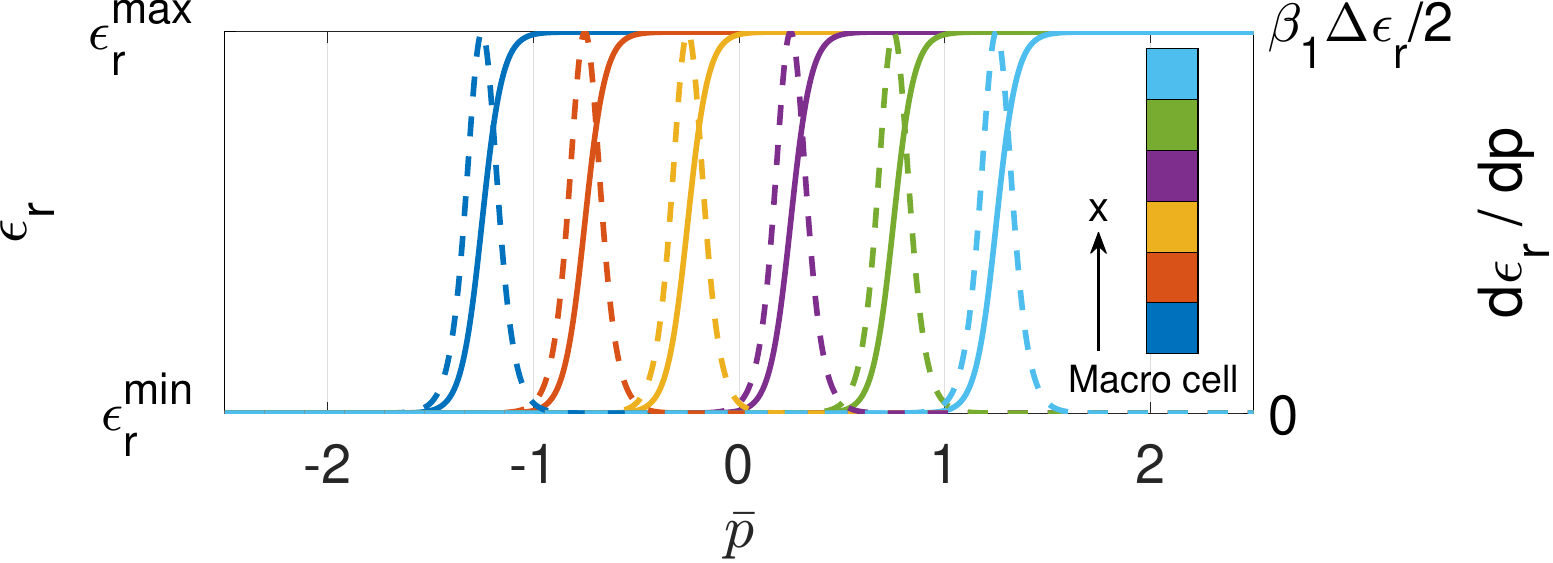}}
		\caption{Mapping function for a macrocell with $N_c = 6$.}
		\label{fig:Mapping_Function_Plot}
	\end{figure}

To ensure a differentiable and physically realizable material parameterization throughout the optimization, the design variables are smoothly mapped to bounded relative permittivity values, as illustrated in Fig.~\ref{fig:Mapping_Function_Plot}. The mapping constrains the relative permittivity to the interval $[\epsilon_r^{\min},\,\epsilon_r^{\max}]$ while providing the continuous first-order derivatives required for the adjoint optimization. Restricting the search space to this structured family of realizable material distributions improves optimization robustness and reduces the occurrence of spurious local minima without significantly degrading the attainable device performance~\cite{Mansouree, ZinLin-inverse, Hammond-Foundary}.
	
The relative permittivity of the $n$-th unit cell is parameterized as
\begin{equation} \label{Relative_Permittivity_Mapping}
	\epsilon_r^n =
	\tanh\!\left(
	\beta_1\bar p_n+\beta_2x_n'+\beta_3
	\right)
	\left(
	\frac{\Delta\epsilon_r}{2}
	\right)
	+\bar\epsilon_r,
\end{equation}
where $\Delta\epsilon_r=\epsilon_r^{\max}-\epsilon_r^{\min}$, and $\bar\epsilon_r = (\epsilon_r^{\max}+\epsilon_r^{\min})/2$. Here, $\bar p_n$ denotes the average value of the design parameter over the macrocell, while $x_n'$ is the local longitudinal coordinate measured within the macrocell. The scalar coefficients $\beta_1$, $\beta_2$, and $\beta_3$ control the slope, longitudinal bias, and offset of the mapping, respectively.
	
	Furthermore, owing to the symmetry of the desired electromagnetic response with respect to the plane $y = 0$, we enforce the structural constraint,
	\begin{equation}
		\begin{split}
			\varepsilon_{r}(x, y) &= \varepsilon_{r}(x, -y), \; \text{for flat metalens,} \\
			\varepsilon_{r}(r, \phi) &= \varepsilon_{r}(r, -\phi), \; \text{for cylindrical metalens,}
		\end{split}
	\end{equation}
	thereby restricting the admissible designs to mirror-symmetric permittivity distributions. Consequently, only the relative permittivity of the left half of the lens is parameterized, with the right half determined by symmetry. This constraint reduces the dimensionality of the optimization problem, remains consistent with the target field symmetry, and improves numerical efficiency without compromising the achievable performance.
	
	For additive manufacturing, a material with relatively low dielectric contrast was chosen to suppress parasitic reflections and improve transmission efficiency \cite{Capasso}. The dielectric properties of the constituent material were characterized by measuring 3D-printed cuboid samples of Polymaker natural PolyLite PLA placed at the center of a WR-28 waveguide section, following the procedure described in \cite{Vasilis-Modulated}. The measurements yielded an average relative permittivity of $\Re\{\epsilon_r\}=2.5$ and a loss tangent of $\tan(\delta)=0.020$ at $30\,\mathrm{GHz}$. Accordingly, the simulations assume $\epsilon_r^{\max}=2.5$ for the 3D-printed material and $\epsilon_r^{\min}=1$ for free-space.
	 
	 During optimization, the mapping parameters are chosen as
	 \begin{equation} \label{Mapping_Parameters_Eq}
	 	\beta_1=10,\qquad
	 	\beta_2=-\frac{\beta_1}{2},\qquad
	 	\beta_3=-\frac{\beta_2(N_c-1)}{2},
	 \end{equation}
	 where $N_c$ is the number of unit cells along the longitudinal direction of each macrocell. These parameters produce a gradual variation of the effective macrocell permittivity, promoting stable convergence of the optimization algorithm. After convergence, the mapping is sharpened by increasing $\beta_1$ from 10 to 1000 while modifying $\beta_2$ and $\beta_3$ according to \eqref{Mapping_Parameters_Eq}. This continuation strategy progressively drives the design toward a binary material distribution suitable for fabrication.
	
	\subsection{Nonlocal Coupling Mechanism}
	The developed formulation is employed to design a coherent-aperture-expanding metalens operating at $30\,\mathrm{GHz}$ for broadside radiation ($\phi_s = 0^\circ$). The metalens has a width of $30\, \lambda_0$ and is composed of three layers of height-modulated dielectrics with a macrocell size of $0.3 \,\lambda_0 \times 0.1\, \lambda_0$. A feeding array antenna with an aperture width of $ W_{\mathrm{in}} = 3.5\, \lambda_0$ and an element spacing $d_e = 0.5 \lambda_0$ is positioned at a distance of $7\lambda_0$ from the metalens. The target output aperture width is specified as $W_{\mathrm{out}} = 30\, \lambda_0$.
	
	\begin{figure}[!t]
		\centering
		
		\begin{subfigure}{0.42\textwidth}
			\includegraphics[width=\linewidth]{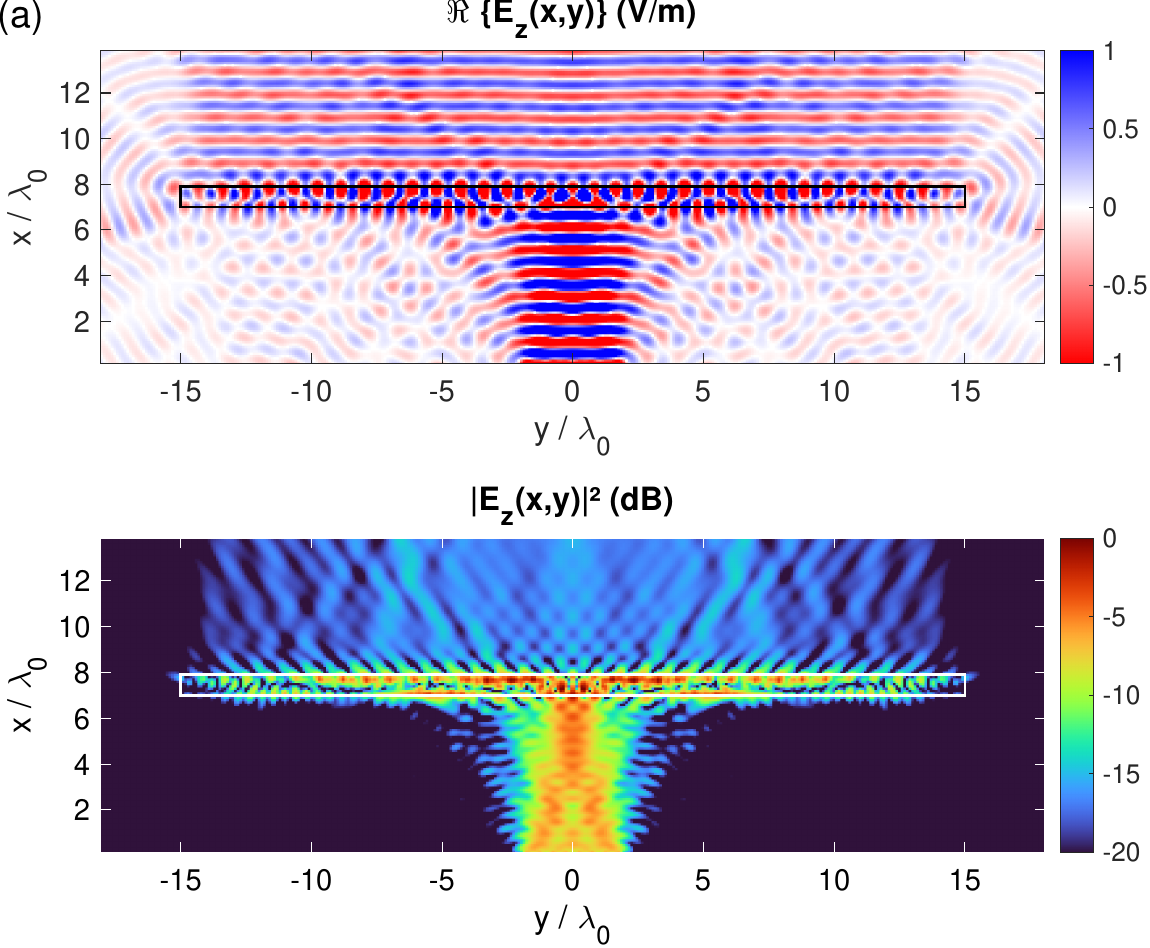}
		\end{subfigure}
		\vspace{0.2cm}
		\begin{subfigure}{0.42\textwidth}
			\includegraphics[width=\linewidth]{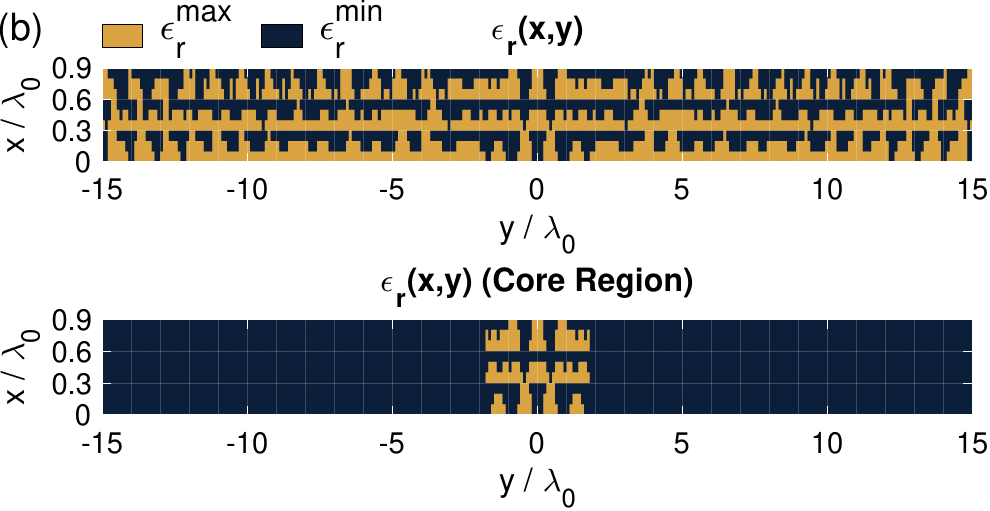}
		\end{subfigure}
		
		\vspace{0.2cm}
		
		\begin{subfigure}{0.42\textwidth}
			\includegraphics[width=\linewidth]{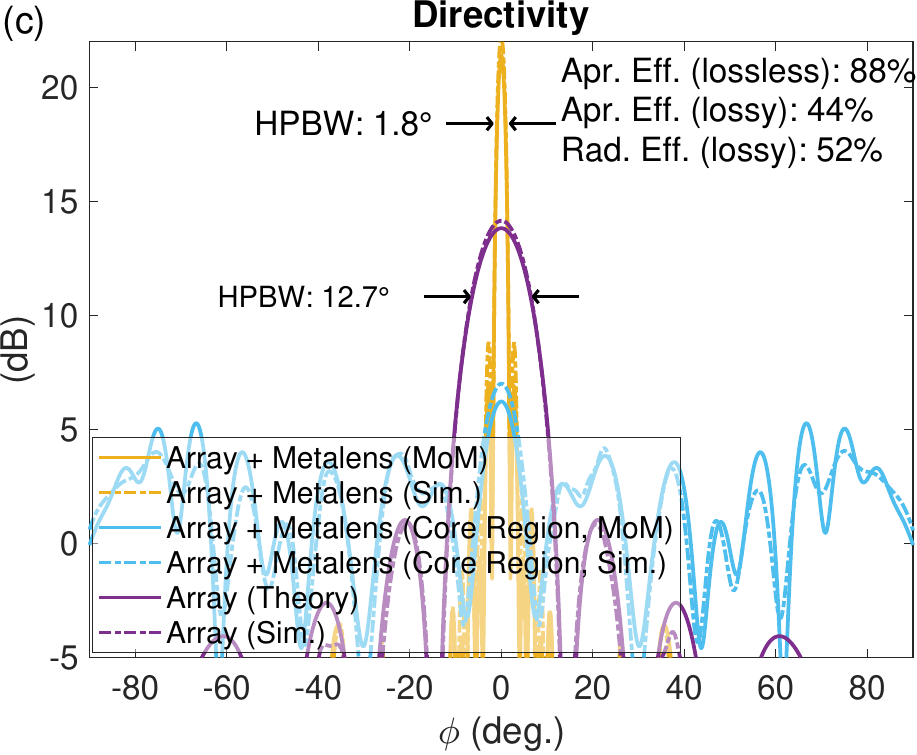}
		\end{subfigure}
		
		\caption{(a) Real part and squared magnitude of the electric field for the beam expander lens at broadside incidence. (b) Relative permittivity profile. (c) Directivity pattern of the feeding array, array with the metalens, and array with the metalens core.} \label{fig:Design}
	\end{figure}
	
	\begin{figure*}[t!]
		\centerline{\includegraphics[width=0.8\textwidth]{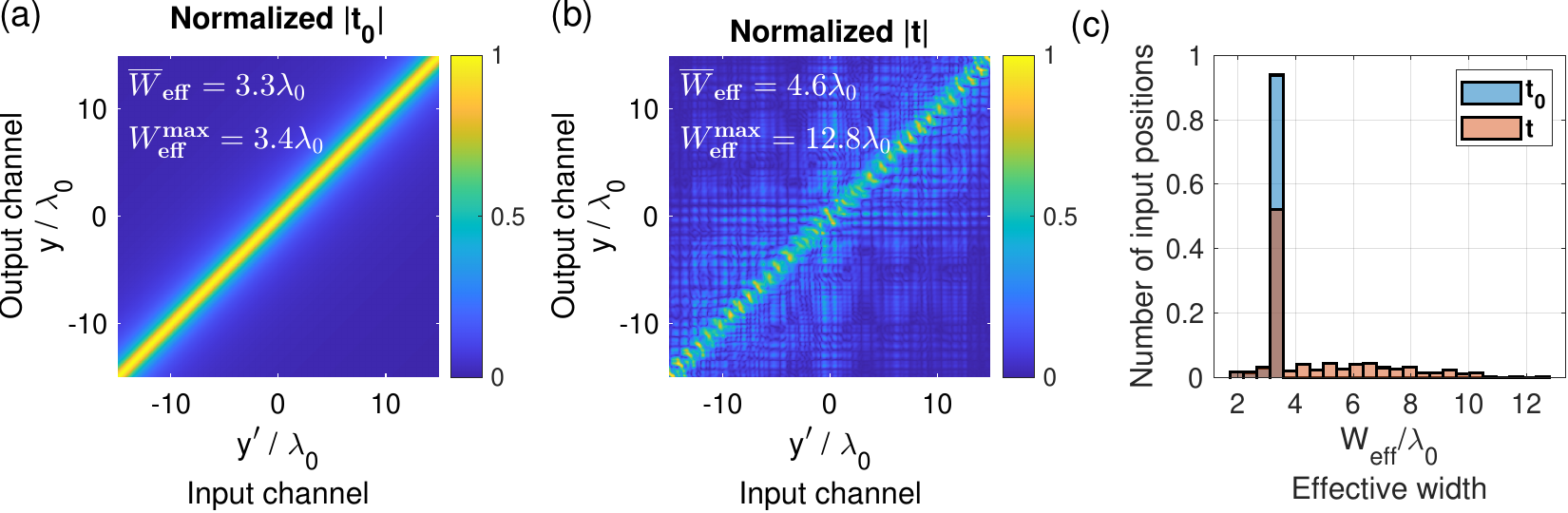}}
		\caption{(a) Normalized free-space transmission matrix. (b) Normalized transmission matrix of the metalens. (c) Histogram of the effective width, compared with that of $\bm{t}_0$.}
		\label{fig:TF}
	\end{figure*}
	
	Fig.~\ref{fig:Design}(a) presents the real part and the intensity of the electric field. The results demonstrate that the incident beam is redistributed by the metalens such that the transmitted field across the output aperture exhibits nearly uniform amplitude and a nearly uniform phase over the prescribed width $W_{\mathrm{out}}$. The transmitted field therefore achieves coherent-aperture expansion across the prescribed output aperture. A closer examination of the optimized permittivity profile in Fig.~\ref{fig:Design}(b) reveals two distinct functional regions. The central region, defined by $|y| \le W_{\mathrm{in}}/2$ and referred to as the core region, primarily converts the incident quasi-planar wavefront into higher-angle propagating components. In contrast, the peripheral region ($|y| > W_{\mathrm{in}}/2$) redirects these components toward broadside, effectively equalizing the phase across the enlarged aperture.
	
	The directivity pattern of the metalens is shown in Fig.~\ref{fig:Design}(c). The narrower main lobe and increased peak directivity indicate successful expansion of the effective coherent aperture. For a lossless material, the calculated aperture efficiency is $88\%$, demonstrating the effectiveness of the proposed approach in expanding the coherent phase region. When material loss is included, the aperture efficiency decreases to $44\%$, while the metalens continues to exhibit the intended directivity enhancement. This reduction highlights the importance of the fill factor and the use of low-loss, 3D-printable materials in realizing high-efficiency metalenses. Furthermore, the directivity is computed for a modified configuration in which only the core region is retained, while the permittivity of the peripheral region is set to unity. In this case, a pronounced increase in radiation within the angular range $10^\circ \le \phi \le 90^\circ$ is observed. This behavior confirms that the core region predominantly couples the incident field into higher-order propagating modes, whereas the peripheral region plays a role in rephasing and redirecting these modes to achieve the desired beam expansion.
	
	The multi-channel transport of the linear system can be described by a spatial transmission matrix of the form
	\begin{equation} \label{Eq:spatial_transmission_matrix}
		\bm{t} = \bm{t}_0 + \bm{G}_1 \bm{L}^{-1} \bm{G}_2,
	\end{equation}
	where $\bm{t}_0$ represents the free-space transmission between the input and output apertures at $x = x^{apr}_{in}$ and $x = x^{apr}_{out}$, respectively, as illustrated in Fig.~\ref{fig:Geometry}(a), and is given by
	\begin{equation}
		[t_0]_{mn} = \frac{k_0}{2j} \frac{x^{apr}_{out}-x^{apr}_{in}}{|\bm{r}_m - \bm{r}_n|} H_{1}^{(2)}\left(k_0 |\bm{r}_m - \bm{r}_n| \right) \Delta y,
	\end{equation}
	where $H_1^{(2)}(\cdot)$ is the first-order Hankel function of the second kind. The matrices $\bm{G}_1$ and $\bm{G}_2$ describe the mutual interactions between the apertures and the lens region and are defined as
	\begin{align}
		[G_1]_{mo} &= \frac{-\omega \mu_0}{4} H_0^{(2)}\left(k_0 |\bm{r}_m - \bm{r}_o| \right) \Delta S_o , \\
		[G_2]_{on} &= \frac{k_0}{2j} \frac{x_{o}-x^{apr}_{in}}{|\bm{r}_o - \bm{r}_n|} H_{1}^{(2)}\left(k_0 |\bm{r}_o - \bm{r}_n| \right) \Delta y ,
	\end{align}
	where $\bm{r}_m = (x^{apr}_{out},~ y_m)$, $\bm{r}_n = (x^{apr}_{in},~ y_n)$, and $\bm{r}_o \in S_v$ denote the position vectors of points on the output aperture, input aperture, and metalens region, respectively. The second term in \eqref{Eq:spatial_transmission_matrix} accounts for field redistribution mediated by the metalens. The transmission matrix maps the electric field on the input aperture to the transmitted field on the output aperture and can therefore be interpreted as the discrete-space impulse response of the system \cite{torfeh}. Unlike free-space propagation, the metalens introduces significant out-of-band components in $\bm{t}$, corresponding to coupling between spatially separated regions of the aperture and providing a direct signature of nonlocal energy transport.
	
	\begin{figure}[th!]
		\centerline{\includegraphics[width=0.32\textwidth]{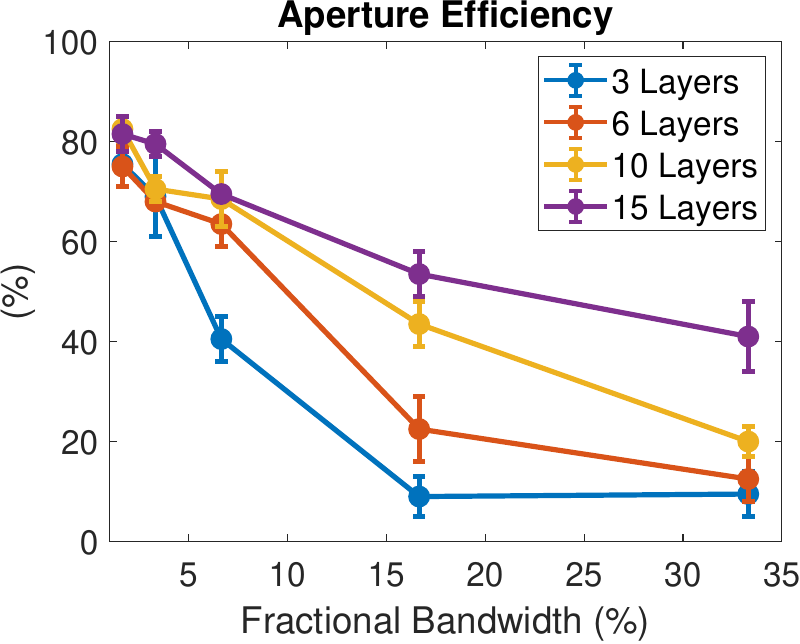}}
		\caption{ Aperture efficiency for increasing fractional bandwidth and number of layers.}
		\label{fig:Percent_Bandwidth}
	\end{figure}
	
	Fig.~\ref{fig:TF}(a) illustrates the free-space transmission matrix $\bm{t}_0$, whose near-diagonal structure reflects the locality of the free-space Green's function between the two aperture planes. In contrast, Fig.~\ref{fig:TF}(b) shows the spatial transmission matrix of the metalens, with the significant out-of-band components appearing relative to $\bm{t}_0$, indicating controlled lateral redistribution of electromagnetic energy across the aperture \cite{Shiyu-thickness}. 	To quantify the degree of lateral spreading, we define the effective spatial transmission width of the $n$th input channel using the inverse participation ratio,
	\begin{equation}
		W_{\mathrm{eff}}
		=
		\frac{\left(\sum\limits_{m} |t_{mn}|^2\right)^2}
		{\sum\limits_{m} |t_{mn}|^4},
	\end{equation}
	which characterizes the effective width over which energy launched from a single input channel is distributed across the output aperture. For a highly localized response, $W_{\mathrm{eff}}$ approaches the free-space value, whereas larger values indicate stronger nonlocal redistribution of energy. We compare this quantity with that of free-space propagation, since the ideal operation of a locally responding lens can be represented as free-space propagation followed by a pointwise phase shift, without mutual coupling. The distribution of the effective width for all the input channels is provided in Fig.~\ref{fig:TF}(c). The average effective transmission width increases from $3.3 \lambda_0$ for the free space propagation to $4.3 \lambda_0$ for the metalens, with the maximum effective spatial transmission increasing from $3.4 \lambda_0$ to $12.8 \lambda_0$, indicating controlled lateral redirection of energy across the aperture. 

\begin{figure}[!t]
		\centering
		
		
		\begin{subfigure}{0.45\textwidth}
			\includegraphics[width=\linewidth]{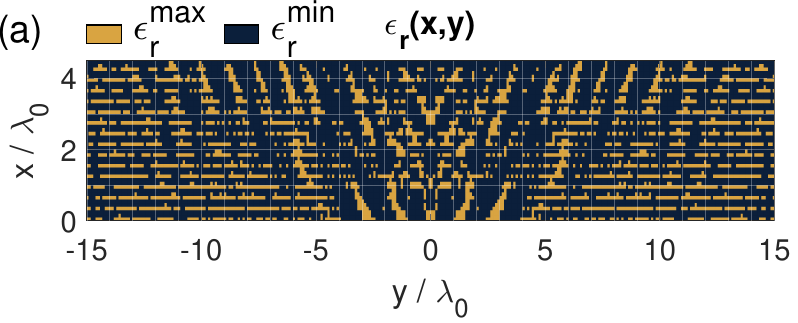}
		\end{subfigure}
		\vspace{0.1cm}
		\begin{subfigure}{0.32\textwidth}
			\includegraphics[width=\linewidth]{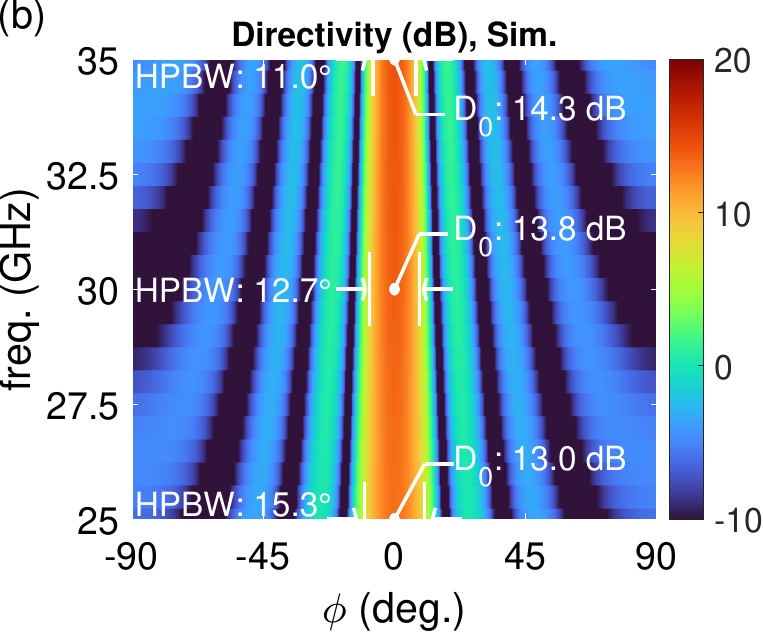}
		\end{subfigure}
		
		\vspace{0.1cm}
		
		\begin{subfigure}{0.32\textwidth}
			\includegraphics[width=\linewidth]{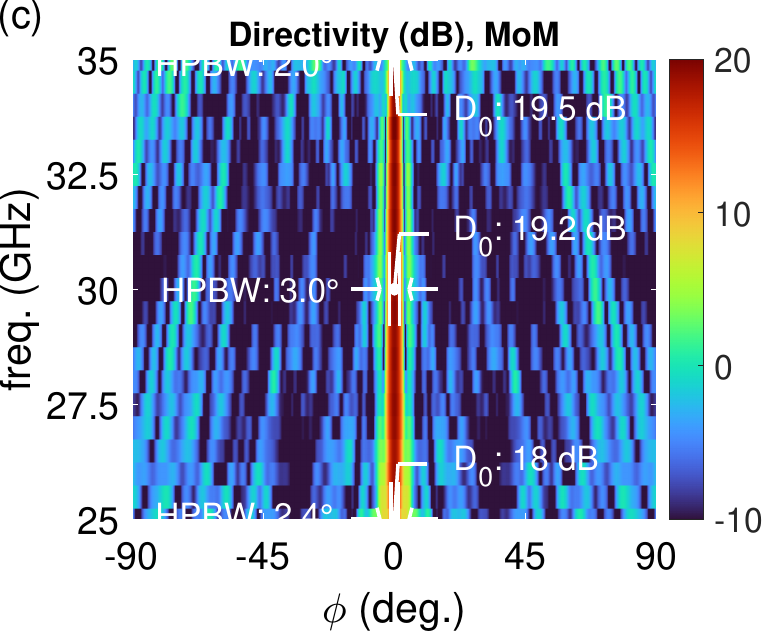}
		\end{subfigure}
		
		\vspace{0.1cm}
		
		\begin{subfigure}{0.32\textwidth}
			\includegraphics[width=\linewidth]{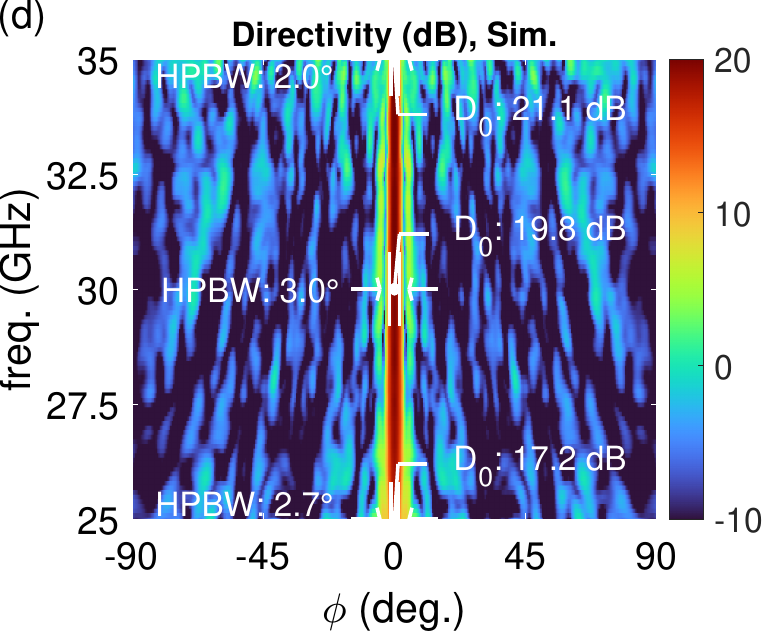}
		\end{subfigure}
		
		\caption{(a) Relative permittivity profile of the metalens. Mirror symmetry with respect to $y = 0$ holds. (b) Directivity pattern of the bare array antenna as a function of frequency. (c) Directivity pattern of the array antenna with the metalens. (d) Simulation results with HFSS. } \label{fig:UWB}
	\end{figure}

	Finally, the proposed design framework can also be extended to broadband operation by optimizing the response simultaneously over frequency. To ensure robust performance over the prescribed range of frequencies, we adopt a worst-case (min-max) optimization formulation,
	\begin{equation} \label{worst_case_optimization_freq}
		\min_{p} \; \max_{\bm{f}} \; F_{f},
	\end{equation}
	where $\bm{f}$ denotes the set of frequencies of interest over which the cost function is evaluated. Figure.~\ref{fig:Percent_Bandwidth} demonstrates that by increasing the number of layers, the additional degrees of freedom enable efficient coherent-aperture expansion over a wide frequency range. For instance, a coherent-aperture-expanding metalens with 15 layers is shown in Fig.~\ref{fig:UWB}(a), which can provide a directivity enhancement of $5.7$--$7.3\, \mathrm{dB}$ over a bandwidth span of $25$--$35\,\mathrm{GHz}$ (Fig.~\ref{fig:UWB}(b,c)). Simulation verification is provided in Fig.~\ref{fig:UWB}(d). These results demonstrate that nonlocal inverse-designed lenses are not inherently narrowband.
	
	\section{Phased-Array Performance Enhancement with Nonlocal Metalenses} \label{Sec:Phased_Array_Performance_Enhancement}
	
	Guided by the channel-capacity bounds, we employ the proposed framework to design metalenses that enhance the directivity of planar phased-array antennas. We assume that the array element spacing and element radiation patterns support grating-lobe-free scanning over the forward half-space, such that the phased array can access the full set of independent propagating angular channels, $N_{\mathrm{in}}$. The framework takes the phased-array aperture width, $W_{\mathrm{in}}$, and the desired output aperture width, $W_{\mathrm{out}}$, as its primary design parameters. These two quantities, together with the channel-capacity constraints, set the achievable directivity enhancement, HPBW, scan resolution, and scan range of the resulting metalens. The MATLAB code implementing the inverse-design pipeline for the nonlocal metalenses reported in this study is available on Zenodo at \cite{Soltani_MATLAB_Code}, and the simulation data generated in this study are available through Figshare at \cite{Soltani_Permittivity_Profiles}.

\begin{figure}[!t]
		\centering
		\begin{tikzpicture}
			\node[anchor=south west, inner sep=0pt] (main) at (0,0) {
				\includegraphics[width=0.38\textwidth]{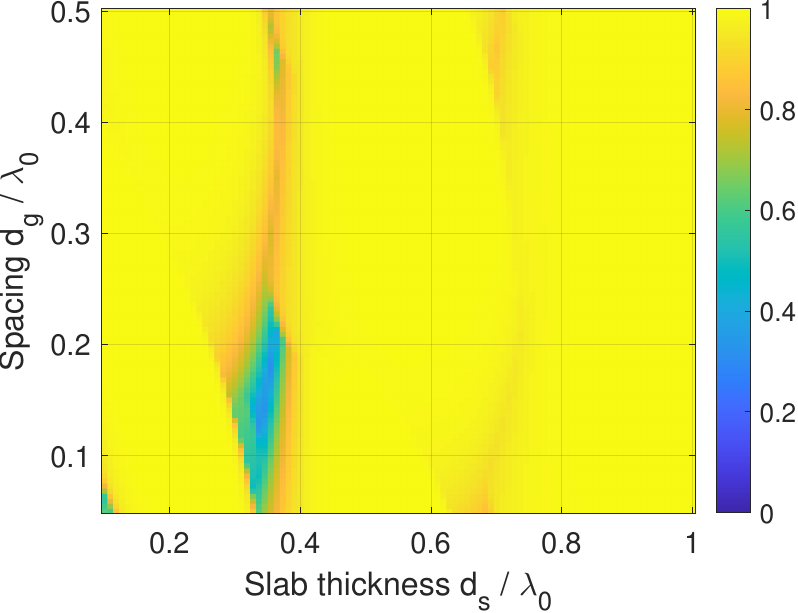}
			};
			
			\node[anchor=north east, inner sep=0pt, draw=black, line width=1pt] at (6.04, 5.20) {
				\includegraphics[width=0.14\textwidth]{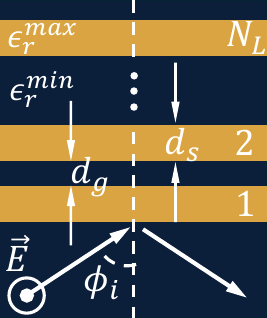}
			};
		\end{tikzpicture}
		\caption{Maximum $|\Gamma|$ for $\phi_{i} \in [0,\, 60]^\circ$, $f \in [29,\, 31] (\mathrm{GHz})$, number of layers $N_L = 10$, $\epsilon_{r}^{\max} = 2.5$, and $\epsilon_{r}^{\min} = 1$ versus slab thickness $d_s$ and spacing $d_g$.}
		\label{fig:MRC}
	\end{figure}

	Owing to the non-convex nature of the inverse-design problem, the final solution generally depends on the initial design. Rather than relying on multiple randomized optimization runs, we initialize the optimization within a physically motivated region of the design space corresponding to broadband, low-reflection multilayer configurations. Specifically, the slab thickness, interlayer spacing, and number of layers are selected to minimize the maximum reflection coefficient over the prescribed frequency and angular ranges while retaining sufficient degrees of freedom to satisfy the design objectives. This consideration becomes particularly important at wide scan angles, where the angle-dependent reflection from the multilayer dielectric can become significant and adversely affect the transmitted field. Consequently, careful selection of the initial multilayer configuration helps avoid starting the optimization from a highly reflective state and provides a favorable design point for subsequent optimization. Figure~\ref{fig:MRC} shows the maximum reflection coefficient of a multilayer structure with $N_L = 10$ interleaved dielectric slabs as a function of the slab thickness and interlayer spacing. The region in which the maximum reflection coefficient is below $0.4$ approximately corresponds to the Bloch-impedance-matched region, $Z_B = Z_i$, where $Z_B$ is the Bloch impedance of the periodic multilayer and $Z_i$ is the impedance of the incident plane wave, thereby maximizing power transmission into the structure over the prescribed operating range. Guided by this analysis, the interlayer spacing and slab thickness are initialized to $d_g = 0.1\lambda_0$ and $d_s = 0.2\lambda_0$, respectively. Accordingly, each macrocell, consisting of three unit cells along the direction of propagation, is initialized with its two bottom unit cells assigned $\epsilon_{r}^{\max}$ and its top unit cell assigned $\epsilon_{r}^{\min}$.
	
	To achieve robust performance over the prescribed scan-angle range, the optimization process is divided into multiple iteration batches. Within each batch, half of the iterations are allocated to sequentially cycling through all scan angles of interest, whereas the remaining half are devoted to a worst-case (min--max) optimization,
	\begin{equation}
		\label{worst_case_optimization_angle}
		\min_{p} \; \max_{\Phi} \; F_{(\phi_{s},\, \phi_{t})},
	\end{equation}
	where the set $\Phi$ contains all the pairs of $(\phi_{s},\, \phi_{t})$ that are allowed according to the channel-capacity bounds. The sequential cycling strategy guarantees that every scan angle contributes to the optimization at regular intervals, allowing the design variables to evolve toward reducing the objective for the entire set. This prevents the optimization from becoming dominated by a small subset of $\Phi$, while other angles---which illuminate different regions of the metalens and require distinct local modifications---remain insufficiently optimized.
	
	\subsection{Scan-Resolution Enhancement via Angular-Channel Compression} \label{SubSec:Scan_Resolution}
		A metalens designed to operate in the angular-channel compression regime concentrates the far-field angular degrees of freedom into a narrower angular sector, thereby enhancing the scan resolution of the phased array, as illustrated in Fig.~\ref{fig:Regimes}(a). For this operating regime, the optimization is performed over the set of input and output scan-angle pairs

\begin{equation}
	\label{Scan_Angle_Pair_Compression}
	\begin{split}
		\Phi =
		\bigg\{
		(\phi_{s,\nu},\,\phi_{t,\nu})
		\,\Big|\,
		&\phi_{s,\nu}
		=
		\sin^{-1}\!\left(
		\frac{\xi_\nu}{W_{\mathrm{in}}}-1
		\right),  \\
		&\phi_{t,\nu}
		=
		\sin^{-1}\!\left(
		\frac{\xi_\nu-W_{\mathrm{in}}}{W_{\mathrm{out}}}
		\right)
		\bigg\},
	\end{split}
\end{equation}
where
\begin{equation} \label{Scan_Angle_Set_Params}
	\begin{split}
		\xi_\nu &=
		\left(\nu-\frac{1-(-1)^{\lfloor N_{\mathrm{in}}\rfloor}}{4}\right)\lambda_0,
		\\
		N_\Phi &=
		\left\lceil\frac{\lfloor N_{\mathrm{in}} \rfloor}{2}\right\rceil,
		\\
		\nu &= 1,\ldots,N_\Phi,
	\end{split}
\end{equation}
	and, $\lceil \cdot \rceil$ and $\lfloor \cdot \rfloor$ denote the ceiling and floor operators, respectively. Note that given the asymptotic nature of \eqref{SBP_Dimension}, $\lfloor N_{\mathrm{in}}\rfloor$ gives a conservative count of the available propagating channels. Additionally, since the metalens response is mirror-symmetric about $y=0$, only half of the angular degrees of freedom need to be considered. Lastly, the ratio $W_{\mathrm{in}} / W_{\mathrm{out}}$ determines the reduction in scan resolution and, consequently, in scan range.
	
\begin{figure*}[!t]
		\centering
		
		\begin{subfigure}{0.83\textwidth}
			\includegraphics[width=\linewidth]{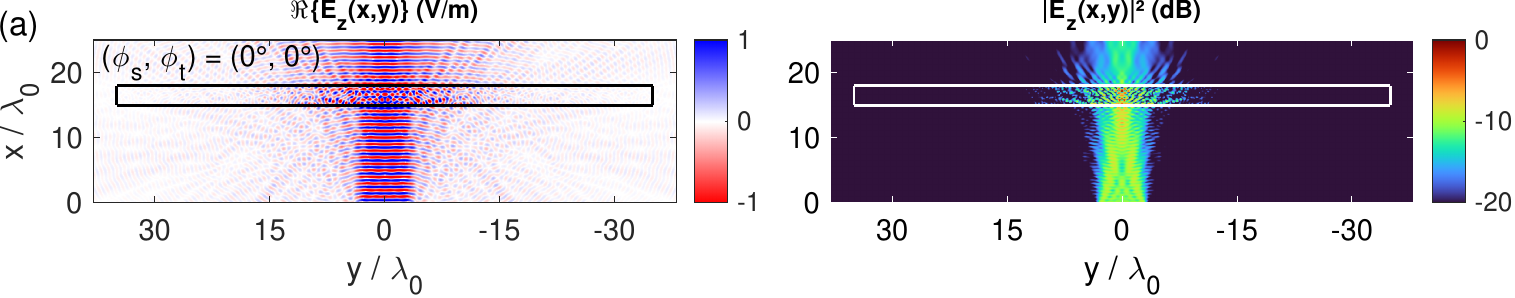}
		\end{subfigure}
		
		\vspace{0.1cm}
		
		\begin{subfigure}{0.83\textwidth}
			\includegraphics[width=\linewidth]{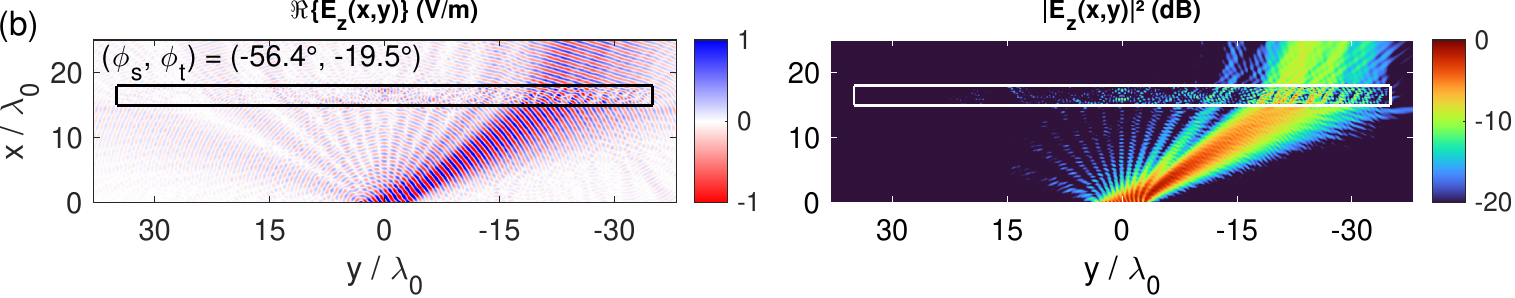}
		\end{subfigure}
		
		\vspace{0.0cm}
		
		\begin{subfigure}{0.83\textwidth}
			\includegraphics[width=\linewidth]{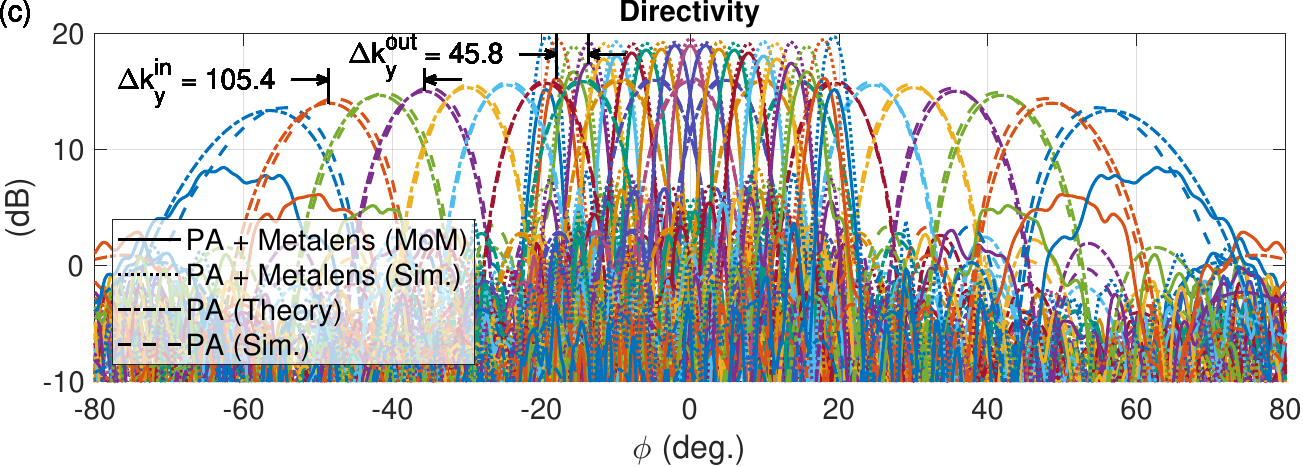}
		\end{subfigure}
		
		\vspace{0.1cm}
		
		\begin{subfigure}{0.83\textwidth}
			\includegraphics[width=\linewidth]{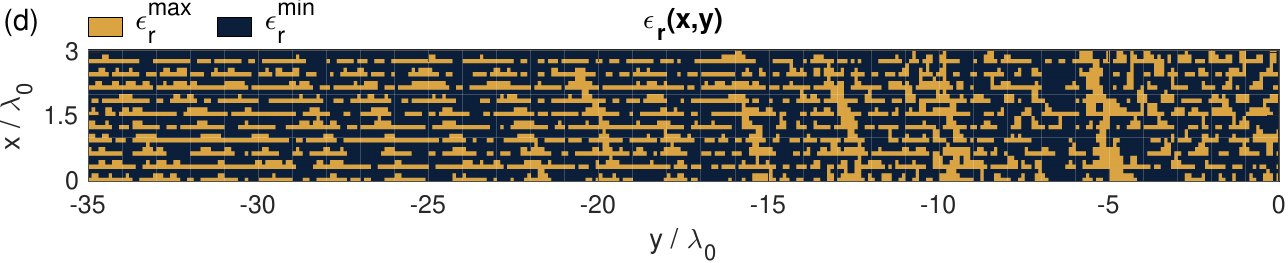}
		\end{subfigure}
		
		\caption{Real part and squared magnitude of the electric field at (a) broadside and (b) $\phi_s = -56.4^\circ$. (c) Directivity pattern without and with the scan-resolution-enhancing flat metalens for $|\phi_s| \le 56.4^\circ$. (d) Relative permittivity profiles of the metalens. The metalens is mirror-symmetric about $y=0$.} \label{fig:fields_flat_compressing}
	\end{figure*}

 Figures~\ref{fig:fields_flat_compressing}(a,b) show the electric-field distributions for a phased array with an aperture width of $6\lambda_0$ and half-wavelength element spacing, integrated with a flat metalens having an output aperture width of $15\lambda_0$, for two representative scan angles. As the scan angle increases, the transmitted wavefront is progressively refracted toward the optical axis, effectively compressing the angular scan range. The directivity patterns in Fig.~\ref{fig:fields_flat_compressing}(c) compare the radiation characteristics of the phased array with and without the metalens. The close agreement between the MoM and HFSS results provides independent full-wave validation of the proposed MoM formulation. The modest discrepancy observed at large scan angles is primarily attributed to the approximation used to represent the radiated fields of the phased array. To demonstrate continuous scanning with this lens, half-channel spacings are used in the channel-counting procedure. The proposed metalens increases the peak directivity by approximately $1.8$--$3.0\,\mathrm{dB}$ while improving the spatial-frequency resolution from $\Delta k_y^{\mathrm{in}} = 105.4\,\mathrm{m}^{-1}$ to $\Delta k_y^{\mathrm{out}} = 45.8\,\mathrm{m}^{-1}$, as indicated in Fig.~\ref{fig:fields_flat_compressing}(c). These values are in close agreement with the theoretical spatial-frequency resolutions of the input and output apertures, $2\pi/W_{\mathrm{in}} = 104.7\,\mathrm{m}^{-1}$ and $2\pi/W_{\mathrm{out}} = 41.9\,\mathrm{m}^{-1}$, respectively. 
	
\begin{figure*}[!t]
		\centering
		\begin{subfigure}{0.83\textwidth}
			\includegraphics[width=\linewidth]{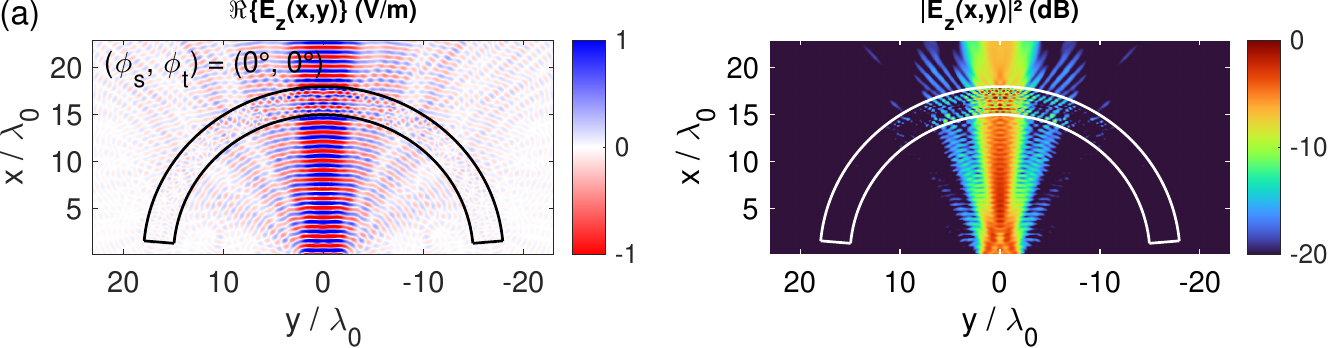}
		\end{subfigure}
		
		\vspace{0.0cm}
		
		\begin{subfigure}{0.83\textwidth}
			\includegraphics[width=\linewidth]{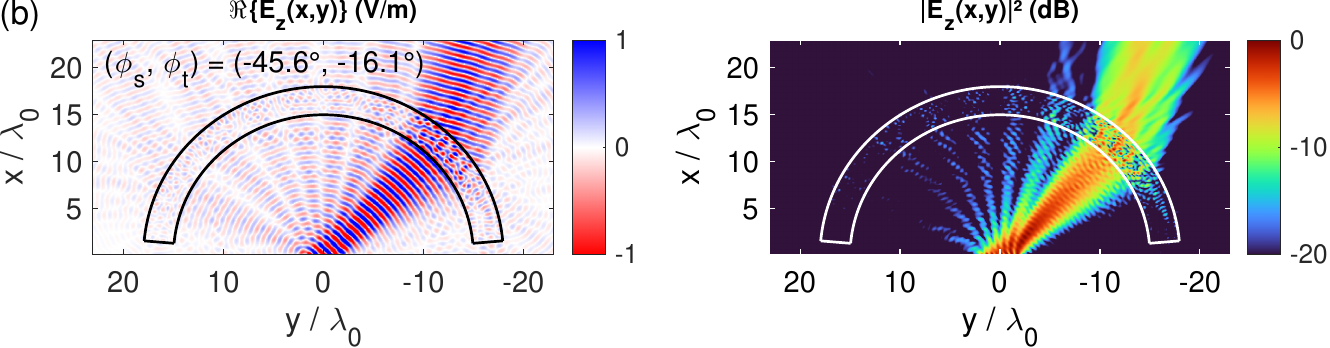}
		\end{subfigure}
		
		\vspace{0.0cm}
		
		\begin{subfigure}{0.83\textwidth}
			\includegraphics[width=\linewidth]{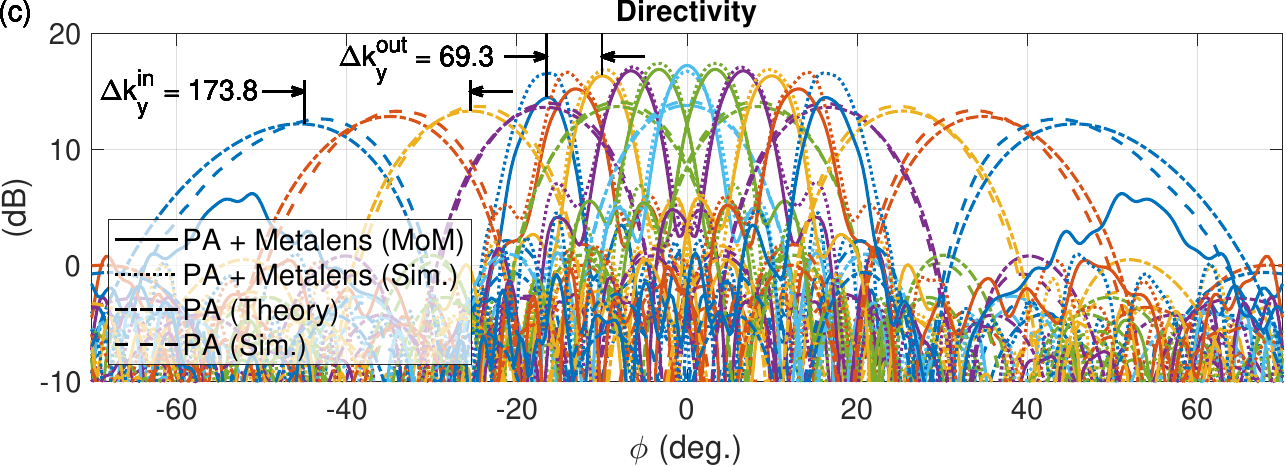}
		\end{subfigure}
		
		\vspace{0.2cm}
		
		\begin{subfigure}{0.83\textwidth}
			\includegraphics[width=\linewidth]{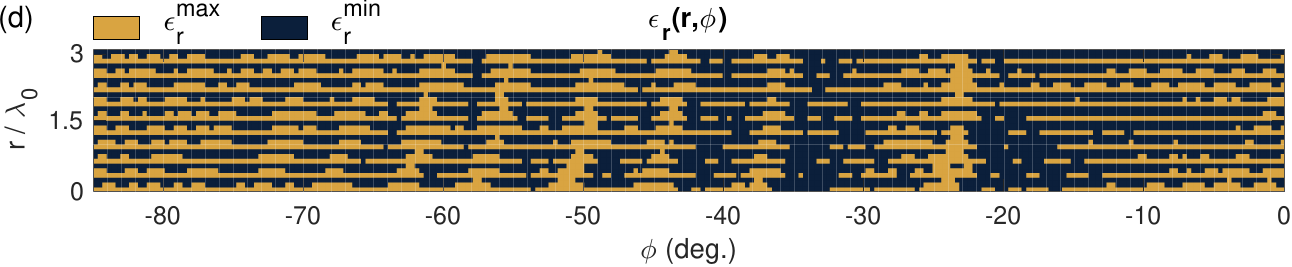}
		\end{subfigure}
		
		\caption{Real part and squared magnitude of the electric field at (a) broadside, and (b) $\phi_s = -45.6^\circ$. (c) Directivity pattern without and with the scan-resolution-enhancing cylindrical metalens for $|\phi_s| \le 45.6^\circ$. (d) Relative permittivity profile of the cylindrical metalens. The metalens is mirror symmetric about $\phi = 0^\circ$.} \label{fig:fields_cylindrical_compressing}
	\end{figure*}

	The bounds derived here are general and therefore also apply to cylindrical metalenses, which offer a more compact alternative to their flat counterparts. As illustrated in Fig.~\ref{fig:fields_cylindrical_compressing}(a,b), the transmitted wavefront is progressively refracted toward broadside as the scan angle increases. For this design, the input and output coherent-aperture widths are set to $W_{\mathrm{in}}=3.5\lambda_0$ and $W_{\mathrm{out}}=9\lambda_0$, respectively, and the corresponding directivity patterns, with and without the metalens, are presented in Fig.~\ref{fig:fields_cylindrical_compressing}(c). As in Fig.~\ref{fig:fields_flat_compressing}(c), half-channel spacings are used in the channel-counting procedure to demonstrate continuous scanning. The HFSS simulations likewise agree closely with the MoM predictions, confirming the robustness of the calculated radiation characteristics for the cylindrical implementation. The metalens provides a peak directivity enhancement of approximately $2.3$--$3.4\,\mathrm{dB}$ across the scan range. As expected, this enhancement is accompanied by angular-channel compression, with the spatial-frequency resolution improving from $\Delta k_y^{\mathrm{in}} = 173.8\,\mathrm{m}^{-1}$ to $\Delta k_y^{\mathrm{out}} = 69.3\,\mathrm{m}^{-1}$. The resulting spatial-frequency resolutions closely match the theoretical values for the input and output apertures, $2\pi/W_{\mathrm{in}} = 179.5\,\mathrm{m}^{-1}$ and $2\pi/W_{\mathrm{out}} = 69.8\,\mathrm{m}^{-1}$, respectively.
	
	A similar angular-channel compression can, in principle, be realized using a conventional hyperbolic phase-delay lens placed in front of a phased array \cite{Tamijani}. In that case, however, each scan angle requires a distinct, scan-dependent excitation profile: the array must synthesize a curved, quasi-spherical wavefront matched to the lens focal geometry to properly illuminate the lens at every pointing direction. In contrast, the proposed metalens is excited using the same standard progressive phase shift between adjacent array elements employed for conventional electronic beam steering, so the array radiates a planar wavefront at every scan angle while still achieving the desired angular-channel compression. Furthermore, because the metalens redistributes energy through engineered nonlocal coupling rather than geometric focusing, the separation between the array and the metalens is largely arbitrary and does not critically affect the radiation characteristics. This contrasts with the conventional hyperbolic lens, where the array must remain near the focal surface, making the array-to-lens distance a critical design parameter that governs the achievable directivity and scanning performance.
		
	\subsection{Directivity Enhancement with Preserved Scan Range via Coherent-Aperture Expansion} \label{SubSec:Scan_Range}
	A metalens operating in the coherent-aperture expansion regime preserves the scan range and scan resolution of the phased arrayƒ, as illustrated in Fig.~\ref{fig:Regimes}(b). Consequently, the set $\Phi$ is defined as 
	\begin{equation}\label{Scan_Angle_Pair_Expansion}
		\Phi =
		\left\{
		\phi_{s,\nu}=\phi_{t,\nu}
		=
		\sin^{-1}\!\left(
		\frac{\xi_\nu}{W_{\mathrm{in}}}-1
		\right)
		\right\},
	\end{equation}
	where $\nu$ and $\xi_\nu$ are defined in \eqref{Scan_Angle_Set_Params}. 
	
\begin{figure*}[!]
		\centering
		
		\begin{subfigure}{0.83\textwidth}
			\includegraphics[width=\linewidth]{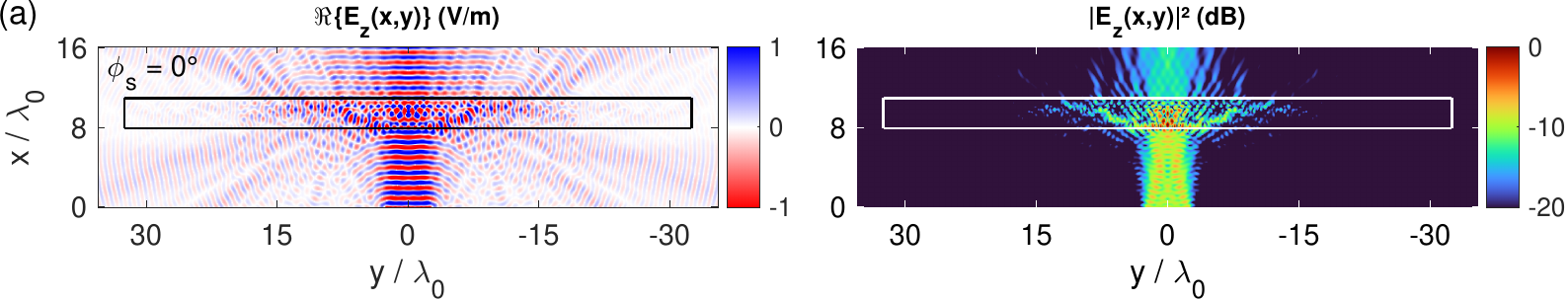}
		\end{subfigure}
		
		\vspace{0.0cm}
		
		\begin{subfigure}{0.83\textwidth}
			\includegraphics[width=\linewidth]{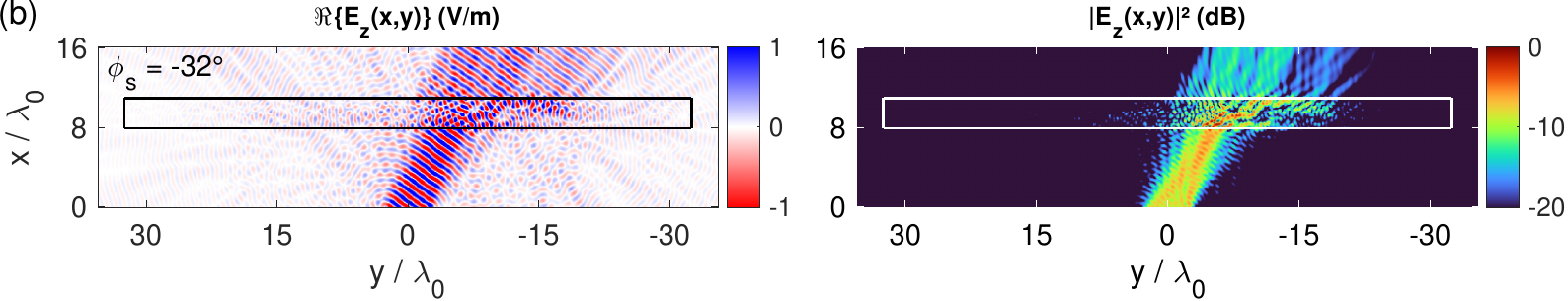}
		\end{subfigure}
		
		\vspace{0.0cm}
		
		\begin{subfigure}{0.83\textwidth}
			\includegraphics[width=\linewidth]{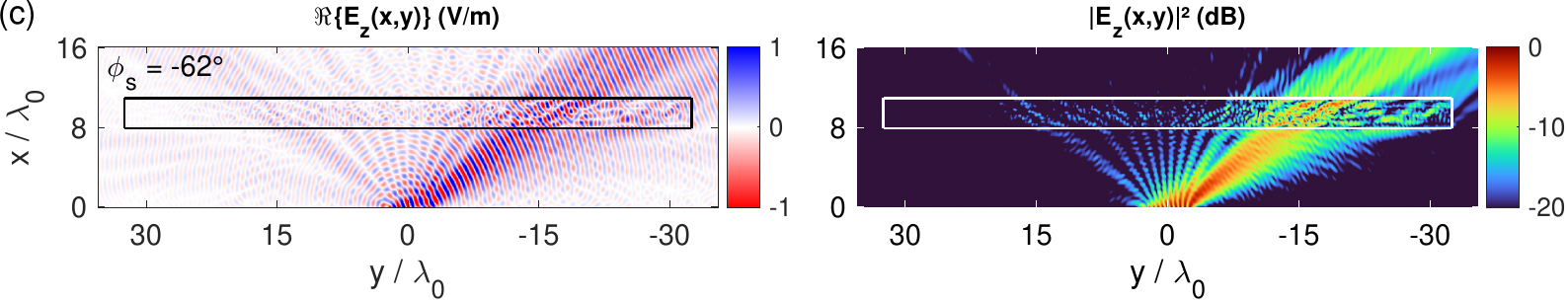}
		\end{subfigure}
		
		%
		
		\vspace{0.0cm}
		
		\begin{subfigure}{0.83\textwidth}
			\includegraphics[width=\linewidth]{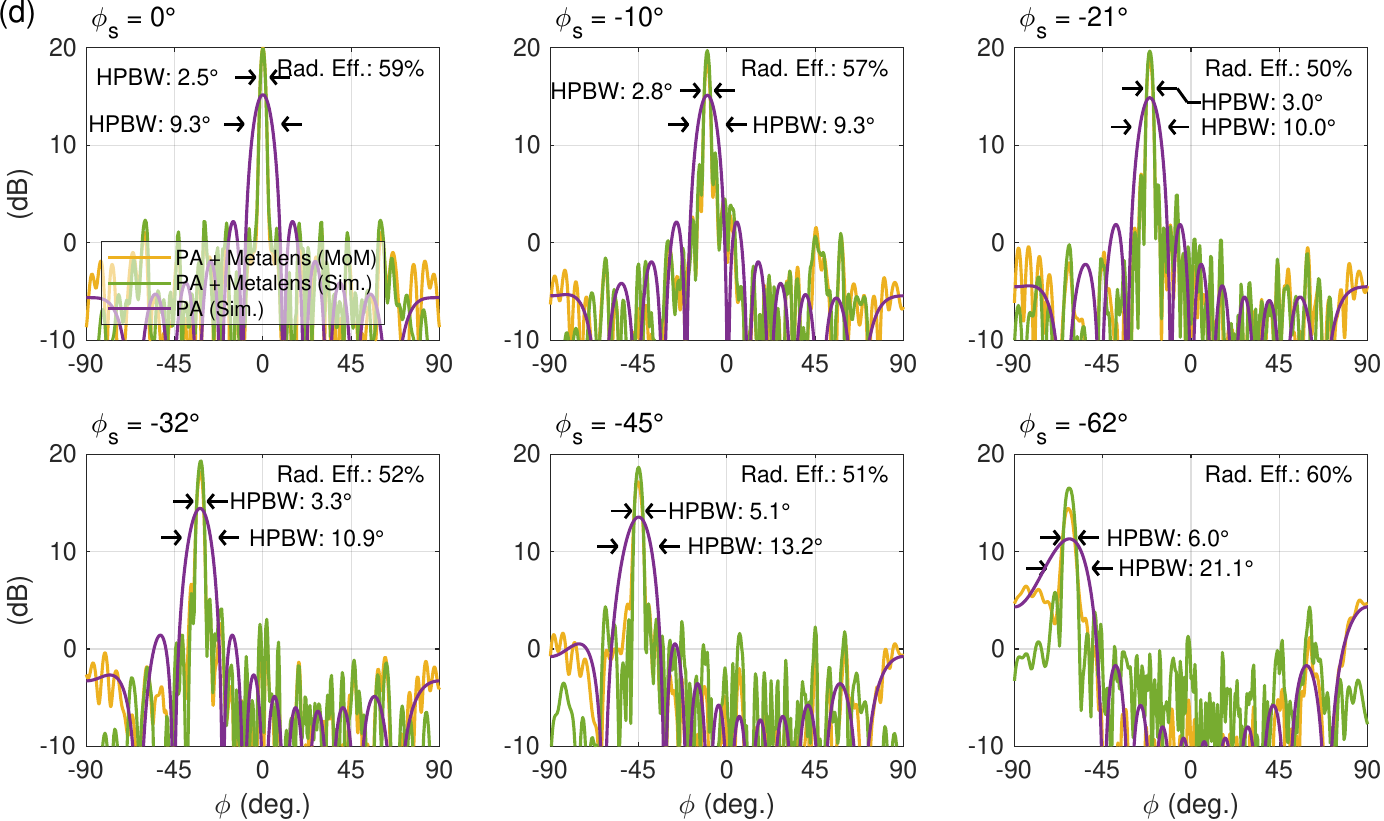}
		\end{subfigure}
		
		\vspace{0.2cm}
		
		\begin{subfigure}{0.83\textwidth}
			\includegraphics[width=\linewidth]{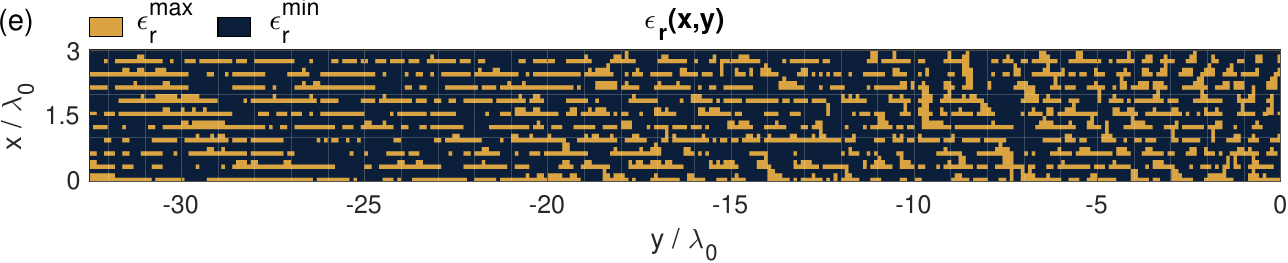}
		\end{subfigure}
		
		\caption{Real part and squared magnitude of the electric field at (a) broadside and at scan angles of (b) $-32^\circ$, and (c) $-62^\circ$. (d) Directivity patterns for $\phi_s = 0^\circ$, $\phi_s = -10^\circ$, $\phi_s = -21^\circ$, $\phi_s = -32^\circ$, $\phi_s = -45^\circ$, and $\phi_s = -62^\circ$. (e) Relative permittivity profiles of the metalens. The metalens is mirror-symmetric about $y=0$.} \label{fig:fields_flat_expanding}
	\end{figure*}

	 For the flat metalens designed for an input aperture of $W_\mathrm{in} = 3.5\lambda_0$ and an output aperture of $W_\mathrm{out} = 20 \lambda_0$, the electric-field distributions for three representative scan angles are shown in Fig.~\ref{fig:fields_flat_expanding}(a--c). In all three cases, the metalens expands the spatial extent of the equi-phase region generated by the phased array while maintaining uniform amplitude. Consistent with the channel-capacity bound derived in Sec.~\ref{Sec:Fundamental_Limits}, the metalens preserves the number of propagating angular channels while redistributing them over a larger coherent aperture, thereby maintaining the scan range and scan resolution despite the increased effective aperture size.
	
		\begin{figure*}[!t]
		\centerline{\includegraphics[width=0.83\textwidth]{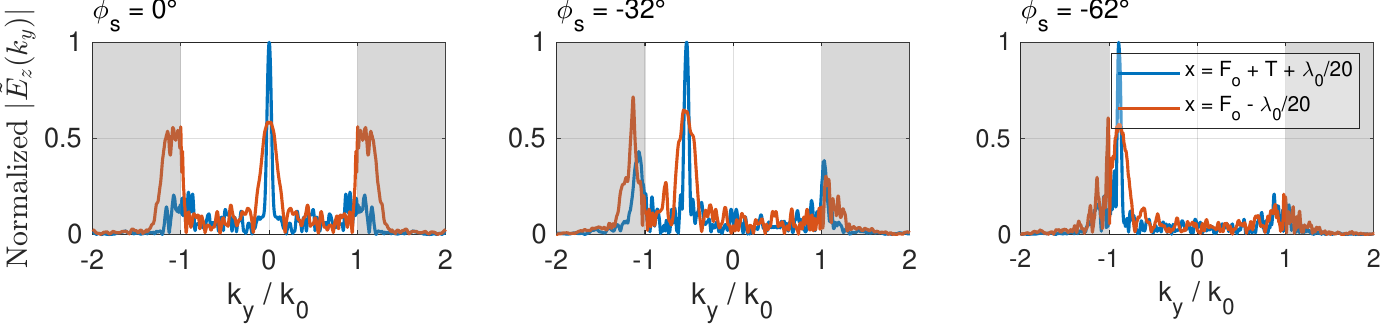}}
		\caption{Spectrum of the electric field on aperture planes located $\lambda_0/20$ from the top and bottom surfaces of the metalens for $\phi_s = 0^\circ$, $\phi_s = -32^\circ$, and $\phi_s = -62^\circ$.}
		\label{fig:fields_flat_expanding_spectrum}
	\end{figure*}
	
	The corresponding directivity patterns are shown in Fig.~\ref{fig:fields_flat_expanding}(d). The directivity enhancement results from the combined effects of a more favorable aperture-amplitude distribution and an enlarged region of phase coherence, which together strengthen constructive interference in the far field. The design was further validated using full-wave simulations in Ansys HFSS, with dielectric losses included. Across the investigated scan angles, a directivity enhancement of $2.5$--$4.8\, \mathrm{dB}$ is achieved while maintaining a radiation efficiency exceeding $50\%$, consistent with the measured loss tangent of the PLA. The relative permittivity profile of the metalens is shown in Fig.~\ref{fig:fields_flat_expanding}(e). Furthermore, the wide-angle beam expansion is facilitated by waves that are evanescent along the propagation direction. Figure~\ref{fig:fields_flat_expanding_spectrum} presents the spatial spectrum of the electric field evaluated on planes located $\lambda_0/20$ above and below the metalens. Within the propagating region, $|k_y|<k_0$, the dominant spectral peak at the output is noticeably narrower than that at the input, indicating an expanded beam in the spatial domain, consistent with the enhanced directivity observed in the far field. In addition, several spectral components with appreciable amplitude appear in the evanescent region, $|k_y|>k_0$. These components are absent in conventional locally periodic phase-delay-profile lenses and facilitate the lateral transport of electromagnetic energy across the aperture, enabling coherent-aperture expansion without creating additional propagating angular channels.  
	
\begin{figure*}[!]
		\centering
		\begin{subfigure}{0.80\textwidth}
			\includegraphics[width=\linewidth]{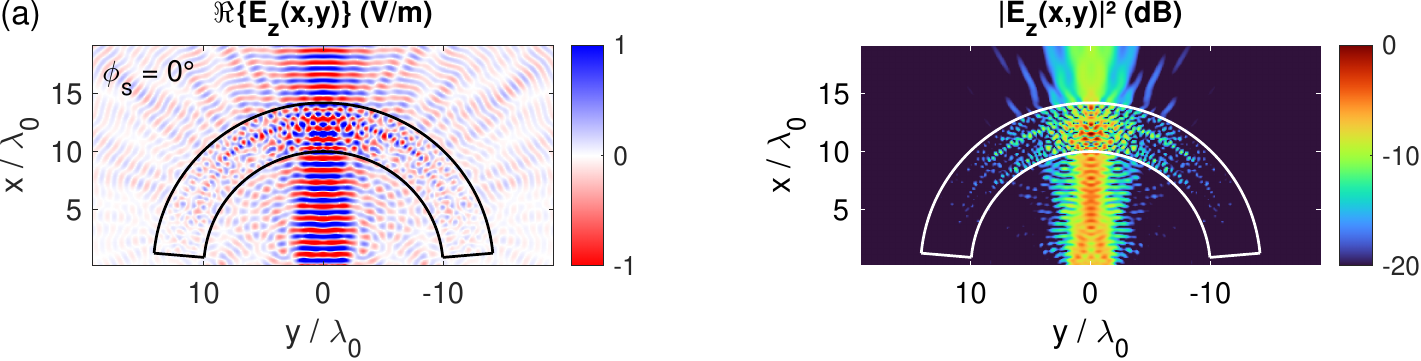}
		\end{subfigure}
		
		\vspace{0.0cm}
		
		\begin{subfigure}{0.80\textwidth}
			\includegraphics[width=\linewidth]{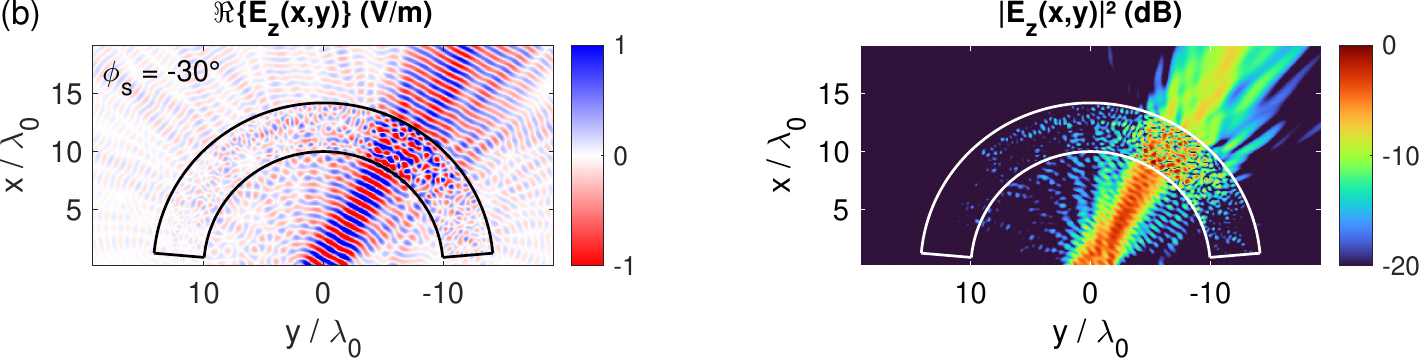}
		\end{subfigure}
		
		\vspace{0.0cm}
		
		\begin{subfigure}{0.80\textwidth}
			\includegraphics[width=\linewidth]{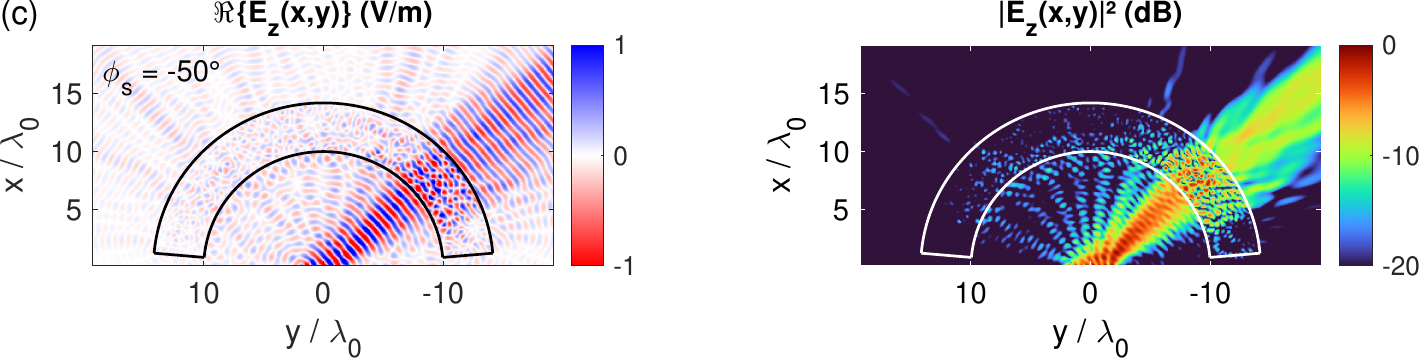}
		\end{subfigure}
		
		\vspace{0.1cm}
		
		\begin{subfigure}{0.80\textwidth}
			\includegraphics[width=\linewidth]{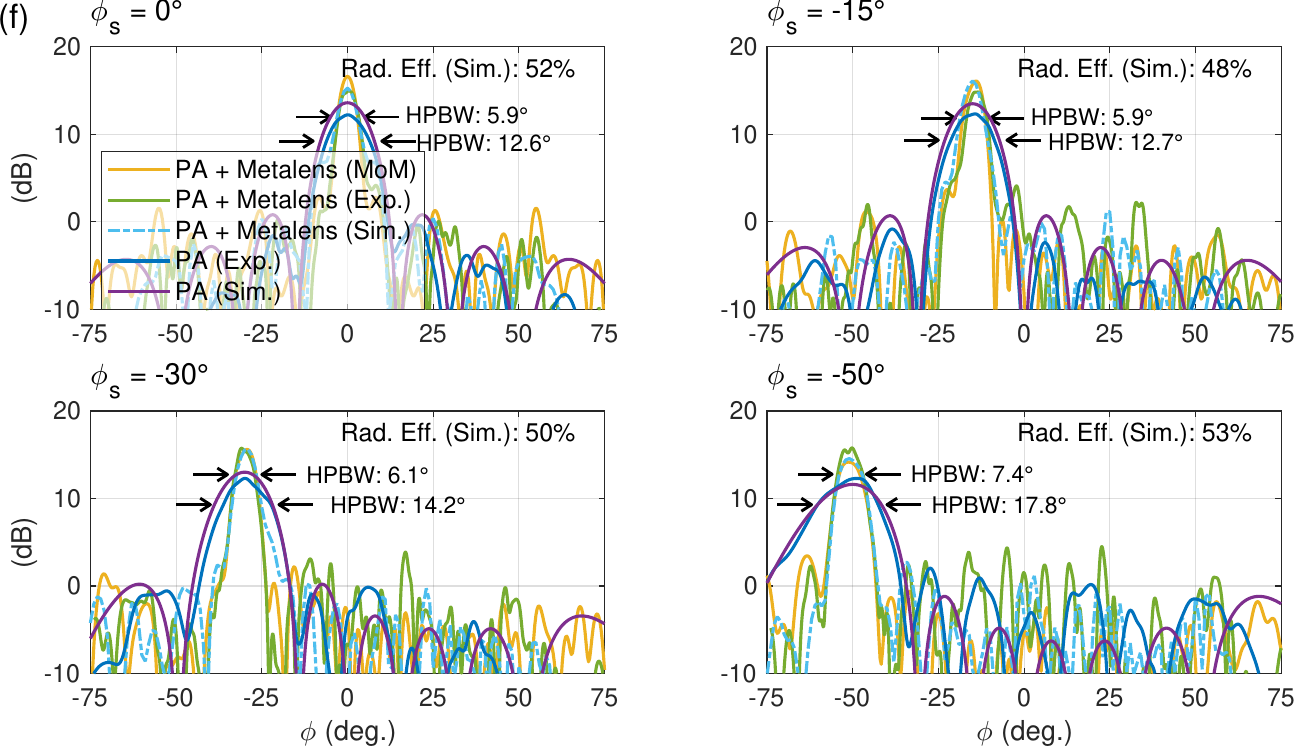}
		\end{subfigure}
		
		\vspace{0.2cm}
		
		\begin{subfigure}{0.80\textwidth}
			\includegraphics[width=\linewidth]{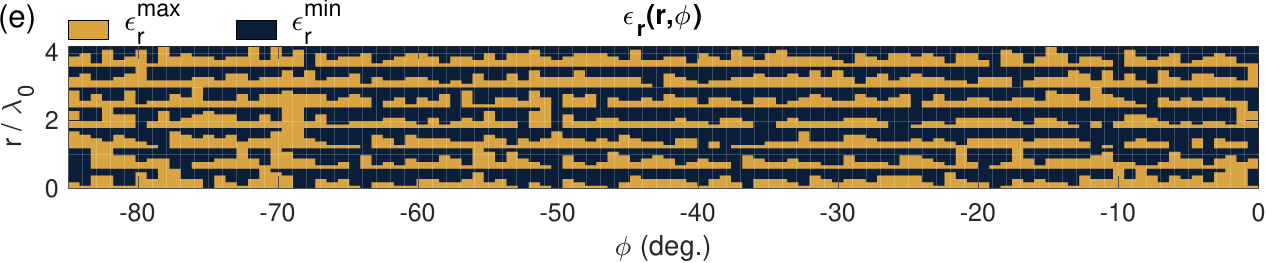}
		\end{subfigure}
		
		\caption{Real part and squared magnitude of the electric field at $29$ GHz for (a) broadside and (b,c) scan angles of $-30^\circ$ and $-50^\circ$, respectively. Directivity patterns for $\phi_s = 0^\circ$, $\phi_s = -10^\circ$, $\phi_s = -21^\circ$, $\phi_s = -32^\circ$, $\phi_s = -45^\circ$, and $\phi_s = -62^\circ$. (e) Relative permittivity profile of the cylindrical metalens. The metalens is mirror symmetric about $\phi = 0^\circ$.} \label{fig:fields_cylindrical_expanding}
	\end{figure*}

	
	An alternative configuration well-suited to phased-array applications is the cylindrical metalens shown in Fig.~\ref{fig:fields_cylindrical_expanding}. In this geometry, the unit cells are implemented as cylindrical sectors with radial length $dr=0.1\lambda_0$ and angular span $d\phi=0.1\lambda_0/(F_o+T_L)$. Each macrocell consists of six unit cells along the radial direction and two unit cells along the azimuthal direction, with a total of $N_L=7$ layers. The cylindrical metalens is designed with $\varphi=85^\circ$ and $F_o=10\lambda_0$. The field distributions are presented in Fig.~\ref{fig:fields_cylindrical_expanding}(a--c). Compared with the planar configuration, the cylindrical geometry can potentially support a larger field of view because the incident wave remains nearly normal to the multilayer structure over the entire scan range. Consequently, it largely suppresses the specular reflections that limit the performance of flat multilayer metalenses at large scan angles.
	
\begin{figure}[!t]
		\centering
		\begin{tikzpicture}
			\node[anchor=south west, inner sep=0pt] (main) at (0,0) {
				\includegraphics[width=0.49\textwidth]{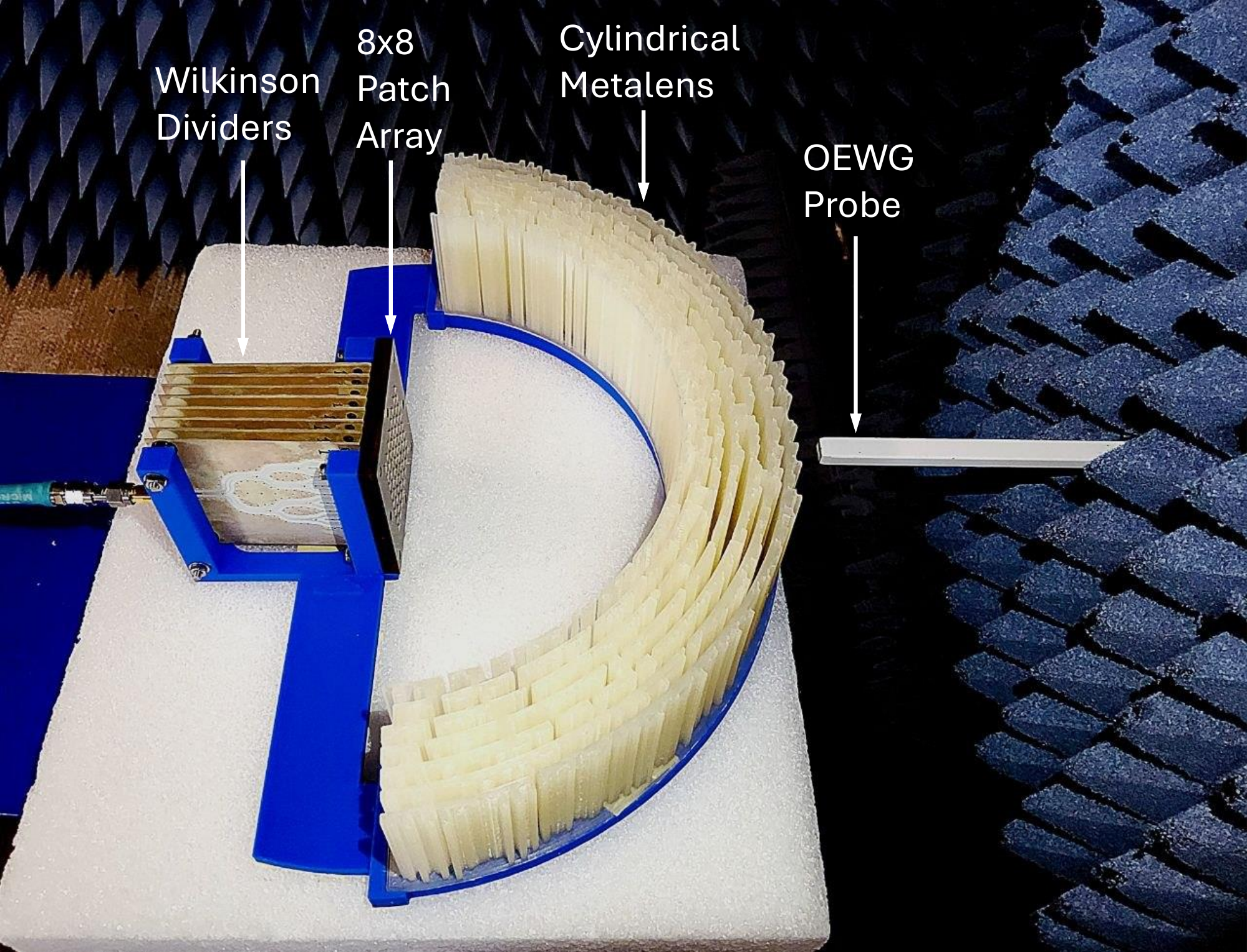}
			};
			
			\node[anchor=north east, inner sep=0pt, draw=black, line width=2pt] at (8.9, 1.85) {
				\includegraphics[width=0.2\textwidth]{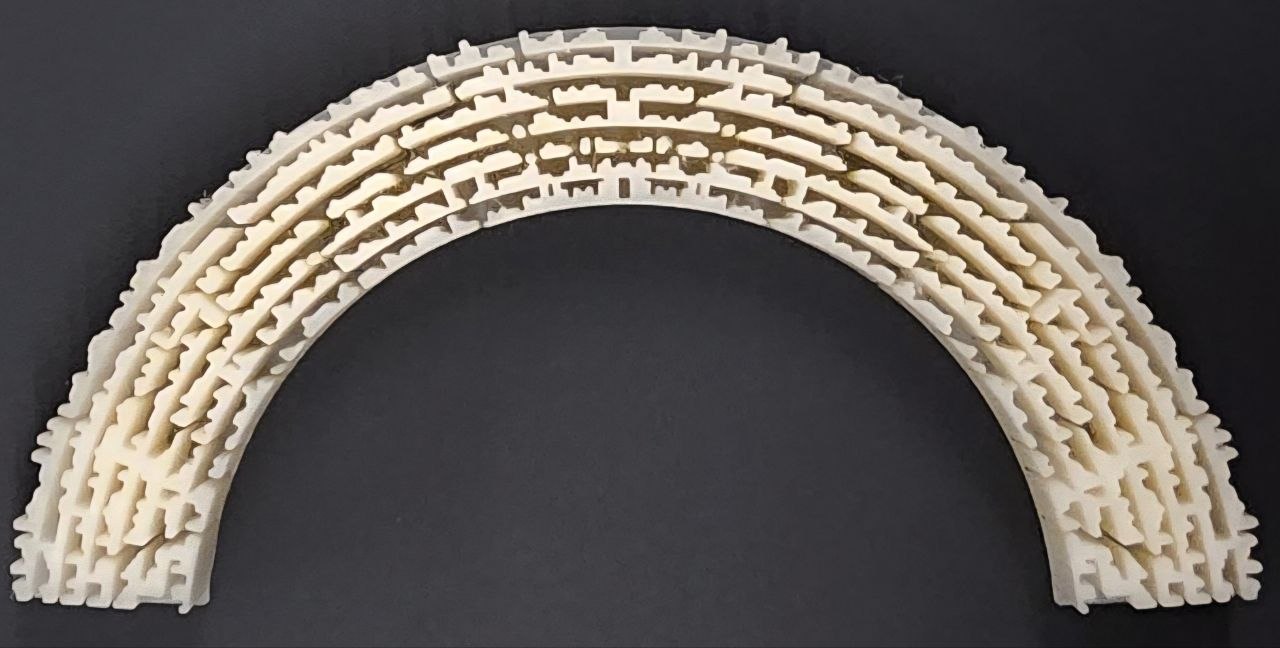}
			};
		\end{tikzpicture}
		\caption{Near-field measurement setup of the array antenna with the cylindrical metalens.  }
		\label{fig:Measurement_Setup}
	\end{figure}

	The proposed design was experimentally validated with the measurement setup shown in Fig.~\ref{fig:Measurement_Setup}, with a photograph of the fabricated 3D-printed metalens provided in the inset. The antenna consists of an $8\times8$ patch array with patch dimensions of $3\,\mathrm{mm}\times 3\,\mathrm{mm}$ and an element spacing of $5\,\mathrm{mm}$ in both directions. The array uses a Rogers 5880 substrate with a thickness of $0.508\,\mathrm{mm}$. For the ground plane, a C10100 ETP copper sheet with a thickness of approximately $6\,\mathrm{mm}$ is used. The antenna substrate is laminated to a copper ground plane using two layers of $0.38\,\mathrm{mm}$-thick Rogers 2929 bondply. The lamination process involves applying approximately $400\,\mathrm{PSI}$ pressure using a $1\,\mathrm{cm}$-thick aluminum pressure frame. The stack is then kept at $245^\circ\mathrm{C}$ for $90$ minutes. The exact temperature profile for 2929 bondply lamination is provided by the manufacturer. Moreover, the ground plane accommodates all 64 straight SMP Mini connectors that excite the patch antennas using a probe with a $0.8\,\mathrm{mm}$ diameter. Eight equal-split 1-to-8 Wilkinson dividers were fabricated to uniformly excite the antenna elements along the z-direction. The dividers were made using microstrip lines and Rogers $\mathrm{RO}\, 4003\mathrm{C}$ substrate with $0.508\,\mathrm{mm}$ thickness. To minimize radiation, co-planar ground plane and guard vias were employed.
	
	The metalens was fabricated using a Bambu Lab P2S 3D printer equipped with a $0.4\,\mathrm{mm}$ nozzle. The printing parameters included a nozzle temperature of $130^\circ\mathrm{C}$, a bed temperature of $60^\circ\mathrm{C}$, an ambient temperature of $35^\circ\mathrm{C}$, $100\%$ infill, and a layer height of $0.2\,\mathrm{mm}$. Because of the printer's build volume of $25.6\,\mathrm{cm}\times25.6\,\mathrm{cm}$, the cylindrical metalens was partitioned into three sections along the air gaps to fit within the build volume for printing, and subsequently reassembled for the measurements. The total printing time was approximately 21 hours.
	
	Near-field measurements were performed using a WR-28 open-ended waveguide (OEWG) probe connected to an E8364B VNA. For each of the eight input ports of the antenna array, the electric field was sampled over a span of $\pm432\,\mathrm{mm}$ with a spatial resolution of $4\,\mathrm{mm}$ at a distance of $157\,\mathrm{mm}$ from the antenna aperture plane. Following the measurements, the channels phases were calibrated to compensate for the insertion-phase mismatch introduced by the Wilkinson power dividers. The calibration was performed using a brute-force optimization in which a unit-magnitude complex coefficient with a phase offset ranging from $-20^\circ$ to $20^\circ$ in $5^\circ$ increments was independently applied to each channel. The combination that provided the closest agreement with the expected radiation pattern was then selected.
	
	 Calibration of the measurements showed the best agreement with the simulations at $29\, \mathrm{GHz}$ rather than the design frequency of $30\, \mathrm{GHz}$, corresponding to an approximately $10\%$ higher relative permittivity of the PLA than assumed in the design. The measured radiation patterns are presented in Fig.~\ref{fig:fields_cylindrical_expanding}(d). Overall, the experimental results agree well with the simulations, demonstrating a peak directivity enhancement of approximately $2.5$--$3.5\, \mathrm{dB}$ while reducing the HPBW by nearly a factor of two. Owing to the mirror symmetry of the cylindrical metalens, comparable performance is expected for positive scan angles as well. Lastly, Fig.~\ref{fig:fields_cylindrical_expanding}(e) presents the relative permittivity profile of the metalens.
	
	Conventional lens-assisted approaches to directivity enhancement, such as Luneburg \cite{Rebeiz, Defocused, Saleem_Lens} and GRIN \cite{LiZhao_Song, LiZhao_2025, Sugimoto} lenses, are fed by a finite number of discrete, non-planar feed elements. By the same rank-composition principle underlying the channel bound derived in this work, a system with $N_{e}$ independent feed ports can synthesize at most $N_{e}$ linearly independent excitation states and therefore cannot access more than $N_{e}$ independent propagating channels, regardless of the lens aperture's own information capacity. Each channel in these systems corresponds to a fixed feed location determined at design time, so the achievable scan resolution is set by the number of feeds and cannot be altered without adding hardware. In the present architecture, the array elements are spaced at the standard half-wavelength interval required for grating-lobe-free scanning, so the number of radiating elements likewise sets the channel budget ($N_{\mathrm{in}} \approx N_{e}$). In both cases, then, the number of independent antenna elements determines the number of accessible propagating channels; what differs is how that fixed budget is accessed—electronically, through phase weighting of a single reconfigurable aperture, versus physically, through selection among discrete, count-limited feed positions.

	\section{Conclusion} \label{Sec:Conclusion}
	This work establishes a general physical framework for understanding the limits of passive linear wavefront engineering in phased-array–lens systems. By formulating the array–lens interaction as a finite-dimensional angular-channel transformation, we show that diffraction and the finite space-bandwidth product of the feeding aperture fundamentally limit the number of available propagating channels. A passive lens cannot create additional channels; it can only redistribute the existing ones. This constraint gives rise to two fundamental operating regimes: angular-channel compression, which improves scan resolution at the expense of scan range, and coherent-aperture expansion, which increases directivity while preserving both scan range and scan resolution. These limits are independent of the specific lens geometry or implementation.
	
	We further show that efficient coherent-aperture expansion requires engineered nonlocal interactions. Through multiple scattering and evanescent-wave-assisted coupling, electromagnetic energy can be redistributed laterally while maintaining phase coherence across an enlarged aperture. Thus, mutual coupling, rather than being merely a parasitic effect, can serve as a fundamental resource for wavefront synthesis.
	
	The derived channel-capacity bounds apply beyond the inverse-designed dielectric metalenses considered here and provide general performance targets for passive array–lens systems, including planar, cylindrical, and other nonlocal wave-transforming structures. By defining the physically attainable design space, the framework can guide inverse-design optimization toward realizable transformations rather than unattainable objectives.
	
	The present analysis is restricted to passive, linear, time-invariant systems and the propagating-wave subspace. Extensions to active and space-time-modulated systems, as well as the joint engineering of spatial, spectral, and polarization channels, may provide new routes toward overcoming these limitations.

	\section*{Acknowledgment}
	The authors acknowledge the computational resources and technical support provided by the Digital Research Alliance of Canada (alliancecan.ca) and Calcul Québec (calculquebec.ca), which enabled the simulations reported in this work. Resource Allocation Project Identifier (RAPI): ctj-430-aa. The authors thank Professor Sean V. Hum for providing access to the near-field antenna testing facility.
	
	\appendices
	
	\section{Analytical Expressions for the MoM Solution} \label{Appendix:Analytical_Expressions}
	
	The matrix $\bm{G}_{vv}$ accounts for the electromagnetic self-interactions and mutual coupling among all discretization cells, thereby capturing multiple scattering within the dielectric structure. Its elements are given by \cite{Harrington}: For $n\neq m$:
	
	\begin{equation}
		[G_{vv}]_{mn} = -\frac{k_0\eta_0\Delta s_{n}}{4} H_0^{(2)}(k_0r_{mn}),
	\end{equation}
	where $r_{mn}$ is the distance between the centers of the source and observation cells, and $\Delta s_{n}$ is the area of the cell $n$. For the self-interaction term, $n=m$, we have
	\begin{equation}
		\begin{split}
			[G_{vv}]_{mn} = &-\frac{k_0\eta_0\Delta s_{n}}{4} \\
			&+j\frac{k_0\eta_0\Delta s_{n}}{2\pi}
			\left[
			\ln\!\left(\frac{k_0r_{\mathrm{eq}}}{2}\right)
			+\gamma
			+S_{\mathrm{shape}}
			\right],
		\end{split}
	\end{equation}
	where $\ln(\cdot)$ denotes the natural logarithm, $r_{\mathrm{eq}}=\sqrt{\Delta s/\pi}$ is the radius of the equivalent circular cell, $\gamma\approx0.57721566$ is Euler's constant, and $S_{\mathrm{shape}}$ is a geometry-dependent shape factor. For a rectangular cell, this factor is given by
	\begin{equation}
		S_{\mathrm{shape}}
		=
		\frac{1}{\Delta s}
		\int_{-dx/2}^{dx/2}
		\int_{-dy/2}^{dy/2}
		\ln\!\left(
		\frac{\sqrt{x^{\prime2}+y^{\prime2}}}
		{r_{\mathrm{eq}}}
		\right)
		dx'\,dy'.
	\end{equation}


\begin{thebibliography}{00}
		
		\bibitem{Heterogeneously_Integrated}
		H. Zhou, R. Jiang, J. Geng, and R. Jin, ``Heterogeneously integrated phased array antennas,'' \emph{Electromagn. Sci.}, vol. 4, pp. 1--27, 2026.
		
		\bibitem{Mailloux}
		R. J. Mailloux, \emph{Phased Array Antenna Handbook}. Norwood, MA, USA: Artech House, 2005.
		
		\bibitem{Rebeiz-1024}
		K. K. W. Low, S. Zihir, T. Kanar, and G. M. Rebeiz, ``A 27--31-GHz 1024-element Ka-band SATCOM phased-array transmitter with 49.5-dBW peak EIRP, 1-dB AR, and $\pm70^\circ$ beam scanning,'' \emph{IEEE Trans. Microw. Theory Techn.}, vol. 70, pp. 1757--1768, 2022.
		
		\bibitem{Quevedo-Teruel-5G}
		O. Quevedo-Teruel, M. Ebrahimpouri, and F. Ghasemifard, ``Lens antennas for 5G communication systems,'' \emph{IEEE Commun. Mag.}, vol. 56, pp. 36--41, 2018.
		
		\bibitem{Hajimiri}
		V. Gurses, S. I. Davis, R. Valivarthi, N. Sinclair, M. Spiropulu, and A. Hajimiri, ``An on-chip phased array for non-classical light,'' \emph{Nat. Commun.}, vol. 16, Art. no. 6849, 2025.
		
		\bibitem{Notaros}
		H. Crawford-Eng, A. Garcia Coleto, B. M. Mazur, \emph{et al.}, ``Reduced-crosstalk antennas for grating-lobe-free and wide-field-of-view integrated optical phased arrays,'' \emph{Nat. Commun.}, vol. 17, Art. no. 3942, 2026.
		
		\bibitem{Shastri-NonlocalFlatOptics}
		K. Shastri and F. Monticone, ``Nonlocal flat optics,'' \emph{Nat. Photon.}, vol. 17, pp. 36--47, 2023.
		
		\bibitem{Vasilis-bandwidth}
		V. G. Ataloglou and G. V. Eleftheriades, ``Low-profile aperiodic metasurfaces for high-efficiency achromatic anomalous reflection over a wide bandwidth,'' \emph{IEEE Trans. Antennas Propag.}, vol. 73, pp. 4703--4715, 2025.
		
		\bibitem{Vasilis-Modulated}
		V. G. Ataloglou and G. V. Eleftheriades, ``Synthesis of modulated dielectric metasurfaces for precise antenna beamforming,'' \emph{Phys. Rev. Appl.}, vol. 19, Art. no. 044033, 2023.
		
		\bibitem{Szymanski-inverse}
		L. Szymanski, G. Gok, and A. Grbic, ``Inverse design of multi-input multi-output 2-D metastructured devices,'' \emph{IEEE Trans. Antennas Propag.}, vol. 70, pp. 3495--3505, 2022.
		
		\bibitem{Estakhri}
		N. Mohammadi Estakhri, B. Edwards, and N. Engheta, ``Inverse-designed metastructures that solve equations,'' \emph{Science}, vol. 363, pp. 1333--1338, 2019.
		
		\bibitem{Momeni-beamforming}
		A. Momeni, \emph{et al.}, ``Reciprocal metasurfaces for on-axis reflective optical computing,'' \emph{IEEE Trans. Antennas Propag.}, vol. 69, pp. 7709--7719, 2021.
		
		\bibitem{Ji-Adjoint}
		S. Ji, S. HuYan, L. Du, X. Xu, and J. Zhao, ``Efficient adjoint-based shape optimization method for the inverse design of microwave components,'' \emph{IEEE Trans. Microw. Theory Techn.}, vol. 73, pp. 494--504, 2025.
		
		\bibitem{Hammond-inverse}
		A. M. Hammond, J. B. Slaby, M. J. Probst, and S. E. Ralph, ``Multi-layer inverse design of vertical grating couplers for high-density, commercial foundry interconnects,'' \emph{Opt. Express}, vol. 30, pp. 31058--31072, 2022.
		
		\bibitem{ZinLin-inverse}
		Z. Lin, C. Roques-Carmes, R. E. Christiansen, M. Soljačić, and S. G. Johnson, ``Computational inverse design for ultra-compact single-piece metalenses free of chromatic and angular aberration,'' \emph{Appl. Phys. Lett.}, vol. 118, Art. no. 041104, 2021.
		
		\bibitem{Zhaoyi-Empowering}
		Z. Li, R. Pestourie, Z. Lin, S. G. Johnson, and F. Capasso, ``Empowering metasurfaces with inverse design: Principles and applications,'' \emph{ACS Photonics}, vol. 9, pp. 2178--2192, 2022.
		
		\bibitem{Zin-Lin-multiwavelength}
		R. E. Christiansen, Z. Lin, C. Roques-Carmes, Y. Salamin, S. E. Kooi, J. D. Joannopoulos, M. Soljačić, and S. G. Johnson, ``Fullwave Maxwell inverse design of axisymmetric, tunable, and multi-scale multi-wavelength metalenses,'' \emph{Opt. Express}, vol. 28, pp. 33854--33868, 2020.
		
		\bibitem{Zhaoyi-Achromat}
		Z. Li, R. Pestourie, J. S. Park, \emph{et al.}, ``Inverse design enables large-scale high-performance meta-optics reshaping virtual reality,'' \emph{Nat. Commun.}, vol. 13, Art. no. 2409, 2022.
		
		\bibitem{Shiyu-metalens}
		S. Li, H. C. Lin, and C. W. Hsu, ``High-efficiency high-numerical-aperture metalens designed by maximizing the efficiency limit,'' \emph{Optica}, vol. 11, pp. 454--459, 2024.
		
		\bibitem{Tamijani}
		A. Abbaspour-Tamijani, L. Zhang, and H. K. Pan, ``Enhancing the directivity of phased-array antennas using lens arrays,'' \emph{Prog. Electromagn. Res. M}, vol. 29, pp. 41--64, 2013.
		
		\bibitem{Landau-Pollak-III}
		H. J. Landau and H. O. Pollak, ``Prolate spheroidal wave functions, Fourier analysis and uncertainty---III: The dimension of the space of essentially time- and band-limited signals,'' \emph{Bell Syst. Tech. J.}, vol. 41, pp. 1295--1336, 1962.
		
		\bibitem{Hansen}
		R. C. Hansen, \emph{Phased Array Antennas}. Hoboken, NJ, USA: Wiley, 2009.
		
		\bibitem{Landau-Pollak-II}
		H. J. Landau and H. O. Pollak, ``Prolate spheroidal wave functions, Fourier analysis and uncertainty---II,'' \emph{Bell Syst. Tech. J.}, vol. 40, pp. 65--84, 1961.
		
		\bibitem{soltani-anisotropic}
		M. Soltani and G. V. Eleftheriades, ``An anisotropic metamaterial cover layer for scan range enhancement of patch-antenna phased arrays in both principal planes,'' \emph{IEEE Open J. Antennas Propag.}, vol. 6, no. 3, pp. 759--773, June 2025.
		
		\bibitem{Harrington}
		R. F. Harrington, \emph{Time-Harmonic Electromagnetic Fields}. New York, NY, USA: McGraw-Hill, 1961.
		
		\bibitem{Mansouree}
		M. Mansouree, A. McClung, S. Samudrala, and A. Arbabi, ``Large-scale parametrized metasurface design using adjoint optimization,'' \emph{ACS Photonics}, vol. 8, pp. 455--463, 2021.
		
		\bibitem{Hammond-Foundary}
		A. M. Hammond, A. Oskooi, S. G. Johnson, and S. E. Ralph, ``Photonic topology optimization with semiconductor-foundry design-rule constraints,'' \emph{Opt. Express}, vol. 29, pp. 23916--23938, 2021.
		
		\bibitem{Capasso}
		L. Sacchi, A. Palmieri, V. Mishra, J.-S. Park, M. Piccardo, and F. Capasso, ``Silica meta-optics: When high performance does not need a high index,'' \emph{Nano Lett.}, vol. 25, pp. 17448--17457, 2025.
		
		\bibitem{torfeh}
		M. Torfeh and A. Arbabi, ``Modeling metasurfaces using discrete-space impulse response technique,'' \emph{ACS Photonics}, vol. 7, pp. 941--950, 2020.
		
		\bibitem{Shiyu-thickness}
		S. Li and C. W. Hsu, ``Thickness bound for nonlocal wide-field-of-view metalenses,'' \emph{Light Sci. Appl.}, vol. 11, Art. no. 338, 2022.
		
		\bibitem{Soltani_Permittivity_Profiles}
		M. Soltani, ``Simulation files to accompany angular-channel capacity and nonlocal wavefront engineering for phased-array–lens systems,'' Figshare, Aug. 2026, doi: \href{https://doi.org/10.6084/m9.figshare.33386542}{10.6084/m9.figshare.33386542}.
		
		\bibitem{Soltani_MATLAB_Code}
		M. Soltani, ``MATLAB code to accompany angular-channel capacity and nonlocal wavefront engineering for phased-array–lens systems,'' Zenodo, Aug. 2026, doi: \href{https://doi.org/10.5281/zenodo.22165063}{10.5281/zenodo.22165063}.
		
		
		\bibitem{Rebeiz}
		B. Schoenlinner, X. Wu, J. P. Ebling, G. V. Eleftheriades, and G. M. Rebeiz, ``Wide-scan spherical-lens antennas for automotive radars,'' \emph{IEEE Trans. Antennas Propag.}, vol. 50, pp. 2166--2175, 2002.
		
		\bibitem{Defocused}
		P. Y. Feng, S. W. Qu, and S. Yang, ``Defocused cylindrical Luneburg lens antennas with phased-array antenna feed,'' \emph{IEEE Trans. Antennas Propag.}, vol. 67, pp. 6008--6016, 2019.
		
		\bibitem{Saleem_Lens}
		M. K. Saleem, H. Vettikaladi, M. A. S. Alkanhal, and M. Himdi, ``Lens antenna for wide-angle beam scanning at 79 GHz for automotive short-range radar applications,'' \emph{IEEE Trans. Antennas Propag.}, vol. 65, pp. 2041--2046, 2017.
		
		\bibitem{LiZhao_2025}
		L.-Z. Song, P.-Y. Qin, Y.-Z. Diao, M. Ansari, J. V. Loesecke, S. Maci, and Y. J. Guo, ``A 3D-printed broadband wide-angle multibeam flat GRIN lens aided by multifocal ray-path analyses,'' \emph{IEEE Trans. Antennas Propag.}, vol. 73, pp. 22--32, 2025.
		
		\bibitem{Sugimoto}
		Y. Sugimoto, T. Tsuchida, S. Takada, K. Sakakibara, T. Narita, and N. Kikuma, ``Homogeneous dielectric multibeam spherical lens antenna with high crossover gain for wide-angle coverage in the 270 GHz band,'' \emph{IEEE Trans. Antennas Propag.}, vol. 73, pp. 6289--6299, 2025.
		
		\bibitem{LiZhao_Song}
		L.-Z. Song, M. Ansari, P.-Y. Qin, S. Maci, J. Du, and Y. J. Guo, ``Two-dimensional wide-angle multibeam flat GRIN lens with a high aperture efficiency,'' \emph{IEEE Trans. Antennas Propag.}, vol. 71, pp. 8018--8029, 2023.	
		
	\end{thebibliography}
\end{document}